\documentclass[a4paper,12pt]{article}
\usepackage{epsfig}
\usepackage{amssymb}
\usepackage{amsfonts}
\usepackage{amsmath}
\usepackage{euscript}
\usepackage{verbatim}
\usepackage{latexsym}
\usepackage{graphicx}
\usepackage{caption, color}
\usepackage{float}
\usepackage{soul}
\usepackage[normalem]{ulem}
\usepackage{amsfonts,amssymb,amsmath, txfonts}
\usepackage{cite, hyperref, cancel}
\usepackage{subcaption}
\usepackage[all]{xy}

\hypersetup{ linktoc=all,
    colorlinks, linkcolor={palatinateblue},
    citecolor={brightpink}, urlcolor={amaranth}
}

\graphicspath{{Images/}}

\definecolor{rosy}{RGB}{230,235,252}
\definecolor{myframetitle}{RGB}{90,89,170}
\definecolor{myblocktitle}{RGB}{140,185,249}
\definecolor{mytitle}{RGB}{10,80,26}

\definecolor{darkgreen}{RGB}{27,130,45}
\definecolor{darkblue}{rgb}{0,0,0.3}
\definecolor{darkred}{rgb}{0.7,0,0}

\definecolor{light gray}{RGB}{220,220,220}
\definecolor{dark purple}{RGB}{108,0,217}
\definecolor{pink}{RGB}{190,20,100}
\definecolor{orang}{RGB}{193,63,0}
\definecolor{green}{RGB}{11,98,17}
\definecolor{darkpink}{RGB}{153,0,76}
\definecolor{bluegreen}{RGB}{0,102,102}
\definecolor{greenlagan}{RGB}{0,102,0}
\definecolor{redgreen}{RGB}{102,102,0}
\definecolor{Redgreen}{RGB}{153,76,0}
\definecolor{vividviolet}{rgb}{0.62, 0.0, 1.0}
\definecolor{amaranth}{rgb}{0.9, 0.17, 0.31}
\definecolor{palatinateblue}{rgb}{0.15, 0.23, 0.89}
\definecolor{brightpink}{rgb}{1.0, 0.0, 0.5}
\definecolor{cornflowerblue}{rgb}{0.39, 0.58, 0.93}
\definecolor{deepcarminepink}{rgb}{0.94, 0.19, 0.22}
\definecolor{radicalred}{rgb}{1.0, 0.21, 0.37}

\usepackage{listings}

\newif\ifdtup

\def\lcdm{$\Lambda$CDM}

\catcode`\@=11

\@addtoreset{equation}{section}

\def\@normalsize{\@setsize\normalsize{15pt}\xiipt\@xiipt
\abovedisplayskip 14pt plus3pt minus3pt%
\belowdisplayskip \abovedisplayskip
\abovedisplayshortskip \z@ plus3pt%
\belowdisplayshortskip 7pt plus3.5pt minus0pt}

\def\small{\@setsize\small{13.6pt}\xipt\@xipt
\abovedisplayskip 13pt plus3pt minus3pt%
\belowdisplayskip \abovedisplayskip
\abovedisplayshortskip \z@ plus3pt%
\belowdisplayshortskip 7pt plus3.5pt minus0pt
\def\@listi{\parsep 4.5pt plus 2pt minus 1pt
     \itemsep \parsep
     \topsep 9pt plus 3pt minus 3pt}}

\relax

\catcode`@=12

\catcode`\@=11

\def\section{\@startsection{section}{1}{\z@}{3.5ex plus 1ex minus
   .2ex}{2.3ex plus .2ex}{\large\bf}}

\def\SymBoxes#1#2#3#4{\newdimen\un@t \un@t#3%
\raisebox{#1}{\rule{#2\un@t}{#4}\hskip-#2\un@t
\@tempdimb\un@t \advance\@tempdimb by-#4\@tempcntb#2\relax%
\@whilenum{\@tempcntb>0}\do{
\rule{#4}{\un@t}\hskip\@tempdimb \advance\@tempcntb by\m@ne}%
\hskip-#2\un@t \rule[\un@t]{#2\un@t}{#4}%
\rule[\un@t]{#4}{#4}\hskip-#4
\rule{#4}{\un@t}}\hskip-#4}                

\begin{document}

\newcommand{\beq}{\begin{equation}}
\newcommand{\eeq}{\end{equation}}
\newcommand{\bea}{\begin{eqnarray}}
\newcommand{\eea}{\end{eqnarray}}
\newcommand{\beas}{\begin{eqnarray*}}
\newcommand{\eeas}{\end{eqnarray*}}
\newcommand{\defi}{\stackrel{\rm def}{=}}
\newcommand{\non}{\nonumber}
\newcommand{\bquo}{\begin{quote}}
\newcommand{\enqu}{\end{quote}}
\renewcommand{\(}{\begin{equation}}
\renewcommand{\)}{\end{equation}}
\def \eqn#1#2{\begin{equation}#2\label{#1}\end{equation}}

\def\e{\epsilon}
\def\IZ{{\mathbb Z}}
\def\IR{{\mathbb R}}
\def\IC{{\mathbb C}}
\def\IQ{{\mathbb Q}}
\def\de{\partial}
\def\Tr{ \hbox{\rm Tr}}
\def\H{ \hbox{\rm H}}
\def\HE{ \hbox{$\rm H^{even}$}}
\def\HO{ \hbox{$\rm H^{odd}$}}
\def\K{ \hbox{\rm K}}
\def\Im{ \hbox{\rm Im}}
\def\Ker{ \hbox{\rm Ker}}
\def\const{\hbox {\rm const.}}
\def\o{\over}
\def\im{\hbox{\rm Im}}
\def\re{\hbox{\rm Re}}
\def\bra{\langle}\def\ket{\rangle}
\def\Arg{\hbox {\rm Arg}}
\def\Re{\hbox {\rm Re}}
\def\Im{\hbox {\rm Im}}
\def\exo{\hbox {\rm exp}}
\def\diag{\hbox{\rm diag}}
\def\longvert{{\rule[-2mm]{0.1mm}{7mm}}\,}
\def\a{\alpha}
\def\dag{{}^{\dagger}}
\def\tq{{\widetilde q}}
\def\p{{}^{\prime}}
\def\W{W}
\def\N{{\cal N}}
\def\hsp{,\hspace{.7cm}}

\def\br{\nonumber}
\def\IZ{{\mathbb Z}}
\def\IR{{\mathbb R}}
\def\IC{{\mathbb C}}
\def\IQ{{\mathbb Q}}
\def\IP{{\mathbb P}}
\def \eqn#1#2{\begin{equation}#2\label{#1}\end{equation}}

\newcommand{\C}{\ensuremath{\mathbb C}}
\newcommand{\Z}{\ensuremath{\mathbb Z}}
\newcommand{\R}{\ensuremath{\mathbb R}}
\newcommand{\rp}{\ensuremath{\mathbb {RP}}}
\newcommand{\cp}{\ensuremath{\mathbb {CP}}}
\newcommand{\vac}{\ensuremath{|0\rangle}}
\newcommand{\vact}{\ensuremath{|00\rangle}                    }
\newcommand{\oc}{\ensuremath{\overline{c}}}
\newcommand{\psizero}{\psi_{0}}
\newcommand{\phizero}{\phi_{0}}
\newcommand{\hzero}{h_{0}}
\newcommand{\psiin}{\psi_{\rh}}
\newcommand{\phiin}{\phi_{\rh}}
\newcommand{\hin}{h_{\rh}}
\newcommand{\rh}{r_{h}}
\newcommand{\rb}{r_{b}}
\newcommand{\psibnd}{\psi_{0}^{b}}
\newcommand{\psibndp}{\psi_{1}^{b}}
\newcommand{\phibnd}{\phi_{0}^{b}}
\newcommand{\phibndp}{\phi_{1}^{b}}
\newcommand{\gbnd}{g_{0}^{b}}
\newcommand{\hbnd}{h_{0}^{b}}
\newcommand{\zh}{z_{h}}
\newcommand{\zb}{z_{b}}
\newcommand{\man}{\mathcal{M}}
\newcommand{\hbr}{\bar{h}}
\newcommand{\tbr}{\bar{t}}

\newcommand\tcr{\textcolor{red}}
\newcommand\tcb{\textcolor{blue}}
\newcommand\tcg{\textcolor{green}}

\newcommand\snote[1]{\textcolor{red}{\bf [Sh:\,#1]}}

\newcommand\ehnote[1]{\textcolor{orang}{\bf [EE:\,#1]}}

\begin{titlepage}
\begin{flushright}
CHEP XXXXX \\
IPM/P-2023/nnnn
\end{flushright}
\bigskip
\def\thefootnote{\fnsymbol{footnote}}

\centerline{\large{\bf{Towards A Realistic Dipole Cosmology: The Dipole $\Lambda$CDM Model}}}


\bigskip
\begin{center}
Ehsan Ebrahimian$^a$\footnote{\texttt{ehsanebrahimianarejan@gmail.com}},\ 
Chethan Krishnan$^b$\footnote{\texttt{chethan.krishnan@gmail.com}}, \ \ Ranjini Mondol$^b$\footnote{\texttt{ranjinim@iisc.ac.in}}, \ \ M. M. Sheikh-Jabbari$^a$\footnote{\texttt{jabbari@theory.ipm.ac.ir}}
\vspace{0.1in}

\end{center}

\renewcommand{\thefootnote}{\arabic{footnote}}

\begin{center}

{$^a$ School of Physics, Institute for Research in Fundamental
Sciences (IPM),\\ P. O. Box 19395-5531, Tehran, Iran}

$^b$ {Center for High Energy Physics, Indian Institute of Science,\\
C. V. Raman Road, Bangalore 560012, India}\\

\end{center}

\noindent
\begin{center} {\bf Abstract} \end{center}
Dipole cosmology is the maximally Copernican generalization of the FLRW paradigm that can incorporate bulk flows in the cosmic fluid. In this paper, we first discuss how multiple fluid components with independent flows can be realized in this set up. This is the necessary step to promote  ``tilted" Bianchi cosmologies to a  viable framework for cosmological model building involving fluid mixtures (as in FLRW). We present a dipole \lcdm\ model which has radiation and matter with independent flows, with (or without) a positive cosmological constant. A remarkable feature of models containing radiation (including dipole $\Lambda$CDM)  is that the \textit{relative} flow between radiation and matter can increase at late times, which can contribute to eg., the CMB dipole. This can happen generically in the space of initial conditions. We discuss the significance of this observation for late time cosmic tensions.


\vspace{1.6 cm}
\vfill

\end{titlepage}

\setcounter{footnote}{0}

\tableofcontents 

\setcounter{footnote}{0}


\section{Introduction and Motivation }

The tension between early and late time observations of the Universe has given rise to what has been called a ``crisis" in modern cosmology \cite{Crisis-1, Crisis-2, Tensions-review, SNOWMASS-2022}. These tensions have only gotten worse over the years, so it may be worthwhile re-examining the basic premises of the present cosmological paradigm. In a recent work \cite{KMS}, we explored the possibility that the Copernican paradigm \cite{Copernicus} is to be relaxed from the usual ``cosmological principle'' to a ``dipole cosmological principle". The idea is to retain the demand that the Universe is homogeneous, but allow the possibility of a flowing cosmic fluid that is consistent with a reduced isotropy group $U(1)$ instead of $SO(3)$. In other words, the isotropy group is broken to the maximal one allowed by a fluid flow.\footnote{This is sometimes referred to as  the assumption of Local Rotational Symmetry (LRS) in the older literature, we will simply refer to it as axial isotropy \cite{KMS}.} The philosophy behind this choice is that priors for the Copernican principle have to be chosen while incorporating the basic observational features of the Universe. If one incorporates only the expansion of the Universe as a prior, one ends up with the current cosmological principle -- at cosmological scales, the background metric should be taken as maximally symmetric on spatial slices (described by an FLRW metric). If one also includes the prior that we see a dipole in the CMB (without necessitating that it is entirely due to our local motions) we end up with the dipole cosmological principle \cite{KMS}. In other words, dipole cosmology is the maximally symmetric generalization of the FLRW paradigm that is compatible with a fluid flow in some direction. 

In pragmatic terms, this resurrects a specific subclass of tilted Bianchi models studied by King and Ellis \cite{King}.\footnote{We prefer the word ``flow" over the word ``tilt" because it emphasizes that the idea is eminently physical. Tilt is not some technical nuance of Bianchi models, but rather an essential characterization of flows. But in deference to the original papers, we will use the two words interchangeably. Note also that in the titled cosmology context the ``tilt'' is a dynamical variable, like the scale factor. 
See \cite{noncomoving} and follow-ups for another approach, where the total stress tensor and metric remain homogeneous and isotropic (FLRW), but the component fluids need not.} See also \cite{vanElst:1995eg, MacCallum, Hewitt:1992sk, Ellis-lectures, Ellis-Maartens, Ellis-Maartens-McCallum--Book, Tsagas} for related discussions. In much of the work on Bianchi models, the flow is set to zero and/or the isotropy group is completely broken. The motivations for the former preference seem purely historical, and must be relaxed if one's goal is to investigate the potential cosmological origins of the CMB dipole. The latter choice is attractive for formal explorations because less symmetry means more generality -- a homogeneous system with a fully broken isotropy group is still governed by a set of ODEs and is therefore relatively tractable. Note however that this is a non-Copernican choice if one's motivations are less formal and more phenomenological. Even when the flow is non-zero, attention has been mainly restricted to formal issues in a single fluid case with a constant equation of state -- see eg., \cite{MacCallum,Ellis-Maartens-McCallum--Book,  Coley-2, Coley-3}. In \cite{KMS, KMS-2} as well as here, our perspective has some important differences compared to these various older works. Let us summarize these.
\begin{itemize}
\item Our goals are decidedly Copernican, which means that we are {\it not} motivated by the idea of breaking as much symmetry as possible in our choice of Bianchi classes. Instead, Copernican philosophy suggests working with the {\em most} symmetric paradigm compatible with the priors. Therefore we are interested in a specific sublcass of the
tilted Bianchi models, those with the maximal leftover isotropy group in the presence of a flow.  The metric is completely fixed -- it is of the axially symmetric Bianchi V/VII$_h$ type \cite{King, Ellis-lectures, Ellis-Maartens-McCallum--Book}. The tilted stress tensor for a general perfect fluid and the equations of motion that generalize the Friedmann equations were presented in \cite{KMS}. 
\item Within this class of models, we are interested in (at least quasi-)realistic phenomenology. Being primarily motivated by cosmic tensions and late time flows, our interest goes beyond simple models with constant equations of state. As a ``poor man's" approach to model building, we considered time-dependent equations of state in \cite{KMS}. This demonstrated that fluid flows can increase at late times in large classes of models including those with late time acceleration. Moreover, we have examples where the fluid flow can grow in intermediate times and go to zero at late times. This generalized the observations in \cite{Coley-1,Coley-2,Coley-3}.
\item The generality of the above observation for large classes of models meant that this should be viewed as a flow instability of the cosmological principle itself under broad conditions \cite{KMS-2}.\footnote{We note that a similar question, with a different viewpoint, was considered in \cite{Coley-2, Coley-3} arguing for  stability of certain perturbations {\em within} certain titled cosmologies.} Explicitly, FLRW cosmologies have fairly generic instabilities against  tilt/fluid flow perturbations. 
In this paper, we will further strengthen the argument that fluid flows are not just a curiosity associated to some specific model within the paradigm -- given the dipole instability of \cite{KMS-2} and the models we discuss here, we could be living in an FLRW universe deformed by a weakly unstable late time flow.
\end{itemize}


In this paper, our goal is to take a few steps towards more realistic phenomenology. We do this by graduating from the ``poor man's" models with time-dependent equations of state to genuine fluid mixtures, like in conventional FLRW model building. Perhaps because of the limited amount of attention that has been given to tilted Bianchi models in general, this does not seem to have been explored much in the literature. One paper where a tilted two-fluid case has been considered is \cite{Coley-1}.\footnote{The paper \cite{Goliath:2000ag} also considers two-fluid case, but with only one tilt and focuses on the late time isotropization of the cosmology.} They work with (a) the Type VI$_0$ Bianchi class and not the axially isotropic Type V/VII$_h$ cases that constitute dipole cosmology; (b) they are interested in formal dynamical systems aspects of the solution space for some specific equations of state,  and (c) they do not consider a cosmological constant.  There are some other mentions of multiple tilted fluid cases in the literature -- most of these do not address this issue in sufficient detail, and some seem incorrect.\footnote{For example, there are papers claiming to construct tilted two-fluid type I Bianchi models, but tilted Type I models are ruled out by theorems in \cite{King}.}

An immediate consequence of allowing mixtures in the dipole cosmology setting is that we can consider fluids with different Equations of State (EoS), as long as we allow them to have {\it independent} flows, all in the same spatial direction so that the isotropy group is not further broken from $U(1)$. We will denote the tilt by a function of cosmic time $\beta (t)$,  with a subscript to distinguish the fluid component when there are multiple fluids with different flows. This is intuitively plausible, and we will show that it is indeed consistent with the symmetries of the set up. This is also in line with the observational hints where reported dipoles are in a broad sense along the CMB dipole direction \cite{SNe-dipole-anomaly,Subir-et-al-QSO, H0-HSA, H0-HSA-2}, see \cite{Beyond-FLRW-review} for a recent review.

The possibility of fluid mixtures enables us to do model-building of the kind familiar in the FLRW paradigm. A particular example that we will elaborate is what we call the {\it dipole \lcdm\ model}. The usual standard model of cosmology, at the level of the background, is described by a 3-component fluid: dark energy (which is modeled by a cosmological constant), matter (dark and baryonic matter which are both modeled as pressureless dust) and radiation (with EoS $w=1/3$). In dipole \lcdm\ model we consider the same mixture of fluids, but in the dipole cosmology setting. Since the cosmological constant is tilt-inert \cite{KMS} our model comes with two independent tilt parameters, one for radiation $\beta_r$ and one for the dust $\beta_m$. We work out the dipole cosmology field equations which are 6 equations for the 6 parameters of the model, $\rho_r, \rho_m, \beta_r, \beta_m$ describing the fluid, as well as the overall expansion rate $H$ and shear $\sigma$ describing the metric. The notations here are natural generalizations of those in \cite{KMS}, and will be elaborated on later. Note that we could have in principle allowed dark matter and ordinary matter to have different flows. This may be necessary for detailed model building and will be discussed more in section \ref{sec:conclusion}. In the present paper, our goal is only to emphasize the model building potential of the setup and observe some basic features.

One such feature we will emphasize in this paper is that the sign of the tilt has dynamical significance. Most discussions of $\beta$ in the literature that we are aware of (including the original King--Ellis paper \cite{King}), consider positive $\beta$. As we will discuss in greater detail later, once the conventions in the metric sector are fixed, both of the signs of $\beta$ for each of the fluid components are physical. We will see classes of cases where the magnitude of $\beta$ can increase at late times, when $\beta$ is negative. In particular, this can happen for radiation.

The possibility of fluid mixtures, together with the choice in sign of $\beta$ for each component of the fluid leads to important phenomenology. We will see that for models involving tilted radiation the {\it relative} flow between dust and radiation can increase at late times. This can happen in the dipole \lcdm\  model, but is  a more general feature of models in which radiation has negative tilt. Indeed, it is a quite generic feature of certain classes of homogeneous perturbations around the FLRW background that break isotropy. 
This observation is  of significance for late time cosmology, because it means that at late times it is quite natural for the cosmic dipole to get a contribution from a homogeneous cosmological flow. Clearly, this is worthy of further study. 

In what follows, we start with a quick recap of dipole cosmology before describing the above features in detail. We will discuss the significance of the sign of tilt first in the context of a dipole $\Lambda$-radiation model, before introducing and elaborating the dipole \lcdm\ model. An Appendix is dedicated to the demonstration that multiple fluids with different tilts are consistent with the dipole cosmology framework. The results of this paper show that model building in dipole cosmology is of theoretical, phenomenological and observational significance and should be developed and studied further.

\section{Review of Dipole Cosmology}\label{sec:2-basic-setup}

Dipole cosmology has two basic ingredients:
\begin{enumerate}
    \item  The anisotropic background metric of the form \cite{KMS, KMS-2}
\bea\label{DipoleMetric}
    ds^{2} = -dt^{2} + a^{2}(t)\left[ e^{4b(t)} dz^{2} +  e^{-2b(t)-2A_{0} z}\big(dx^{2}+dy^{2}\big) \right],
\eea
where $a(t)$ is the overall scale factor, $b(t)$ captures the anisotropy and $A_0\neq 0$ is a constant with dimension of inverse length. It can be set to 1 by a choice of units, but we will keep it for later convenience. 
\item Tilted energy momentum tensor which in the  $(t,z, x,y)$ coordinates  takes the form
\begin{equation}\label{T-tilted}
\begin{split}
    &T^{\mu}{}_{\nu} = T_{\text{iso}}{}^{\mu}{}_{\nu} + T_{\text{tilt}}{}^{\mu}{}_{\nu},\qquad 
T_{\text{iso}}{}^{\mu}{}_{\nu}=\text{diag}(-\rho, p,p,p)\\
&T_{\text{tilt}}{}^{\mu}{}_{\nu}=(\rho+p)\sinh\beta\left(\begin{array}{cccc}
 -\sinh\beta &  ae^{2b}  \cosh \beta & 0 & 0 \\
-\cosh \beta/(ae^{2b}) &  \sinh\beta  & 0 & 0 \\
 0 & 0 & 0 & 0 \\
 0 & 0 & 0 & 0\\
\end{array}
\right)
\end{split}
\end{equation}
where $T_{\text{iso}}{}^{\mu}{}_{\nu}$ is the energy momentum tensor of a usual isotropic perfect fluid and $\rho$ and $p$ denote its  energy density  and pressure. 
\end{enumerate}
Some comments are in order:
\begin{itemize}
    \item The tilt parameter $\beta$ is the rapidity factor for the cosmic fluid from the viewpoint of the observer comoving with the metric \eqref{DipoleMetric} \cite{King}. That is, as is also clearly seen from $T_{\text{tilt}}{}^{\mu}{}_{\nu}$, the observer at constant  cosmic time $t$ slice will see a bulk flow \cite{Bulk-flow} in the anisotropy direction $z$. 
    \item $T_{\text{tilt}}{}^{\mu}{}_{\nu}$ is traceless and therefore the trace of $T^{\mu}{}_{\nu}$ is independent of tilt and is equal to $-\rho+3p$. This may be understood noting that trace of $T$ is a scalar and does not change under boosts.
    \item The tilt $\beta$ drops out when $p=-\rho$, i.e. for a cosmological constant. 
    \item Parity along the tilt direction $z$ is broken once we choose a sign convention (say positive) for $A_0$. 
    \item With the choice of sign of $A_0$, the sign of $\beta$ acquires physical significance. 
\end{itemize}
Later in this section we present the expressions and equations for a single fluid -- the generalization to multiple fluids simply amounts to considering a multi-component stress tensor with corresponding $\rho_i, p_i$ and $\beta_i$ as we elaborate below. This last statement is intuitive when the flows are all in the same spatial direction, see Appendix \ref{Appendix:consistency} for more details.

For a generic $n$ component fluid we have $3n+2$ dynamical variables, $\rho_i, p_i, \beta_i; a, b,\ i=1,2,\cdots n$. Typically, $p_i=w_i \rho_i$ for constant EoS $w_i$ (which are model parameters). Therefore, we have $2n+2$ variables. For a single fluid of given EoS, that is $4$ parameters and for the two fluid case it is 6. Of course cosmological constant $\Lambda$ can also be added to the system without changing number of dynamical fields. For $n$ \textit{non-interacting} fluid components, 
\begin{equation}\label{T-sum}
T^{\mu}{}_{\nu} = \sum_{i=1}^n (T^{\mu}{}_{\nu}){}_i, 
\end{equation} where $(T^{\mu}{}_{\nu}){}_i$ has the same expression as  \eqref{T-tilted} with $\rho=\rho_i,\ p=p_i,\ \beta=\beta_i$. See appendix \ref{Appendix:consistency} for more details. 

Instead of $a,b$ it is sometimes convenient to work with  the Hubble expansion rate $H(t)$ and the cosmic shear $\sigma (t)$
\begin{equation}\label{H-sigma}
H:=\frac{\dot{a}}{a},\qquad \sigma:=3\dot{b} 
\end{equation}
where \textit{dot} stands for derivative with respect to $t$. When $\sigma=0$ the metric reduces to an open FLRW universe. 

The dynamics of these parameters are governed by Einstein equations, 
\bea\label{cosmoC}
R_{\mu\nu}-\frac12 Rg_{\mu\nu}= T_{\mu\nu}, 
\eea
where we have set the units such that $8\pi G$ is equal to 1. While covariant conservation of the total energy momentum tensor follows from consistency of Einstein equations, for non-interacting fluids, we have  the continuity equations for each component separately, $\nabla_\mu  (T^{\mu}{}_{\nu}){}_i=0$. Einstein equations for a generic multi-component fluid has been worked out in appendix \ref{Appendix:consistency} and here we present them for the single fluid case \cite{KMS}:
\begin{subequations}\label{EoM-H-sigma}
\begin{align}
H^2-\frac19\sigma^2-\frac{A_0^2}{a^2} e^{-4b}&=\frac{\rho}{3}+\frac13 (\rho+p)\sinh^2\beta
\label{EoM-H-sigma-c}\\
 \sigma &=\frac{1}{4A_0} a e^{2b}(\rho+p)\sinh2\beta,
\label{EoM-sigma}\\
\dot{\rho}+3H(\rho+p)&=-(\rho+p)\tanh\beta(\dot{\beta}-\frac{2A_0}{a} e^{-2b}) \label{Con1-1} \\
\dot{p}+H(\rho+p)&= -(\rho+p)\left( \frac23\sigma+\dot{\beta}\coth{\beta}\right). \label{Con2-1}
\end{align}
\end{subequations}
We also note that 
\begin{equation}\label{EoM-H-sigma-sigma-dot}
\begin{split}
\dot{\sigma}+3H\sigma &=(\rho+p )\sinh^2\beta \\
\frac{\ddot{a}}{a}=   \dot{H}+H^2 &=-\frac16{(\rho+3p)}- \frac13(\rho+p)\sinh^2\beta-\frac29 \sigma^2
\end{split}\end{equation}
which are of course not independent equations from those in \eqref{EoM-H-sigma}.
Note that \eqref{EoM-sigma}  clearly shows the correlation between the sign choices in $\beta, A_0$ and  $\sigma$.

The above four equations \eqref{EoM-H-sigma} are the extensions of Friedmann equations to a single fluid dipole cosmology, and should be supplemented (as in the usual FLRW setting) by an equation of state to provide a complete set of equations. 
Note that the cosmological constant $\Lambda$ is a fluid with EoS $w=-1$ and is tilt-inert, and can be treated in this framework by a trivial modification of the above equations, see \cite{KMS}. We shall return to this below.
As discussed in \cite{KMS, KMS-2} the above equations can allow for growth of tilt $\beta$ while the shear goes to zero  irrespective of the presence or absence of the cosmological constant. In particular, it was noted in \cite{KMS-2} that this signals an instability in the FLRW paradigm towards tilt deformations. We note that if the null energy condition holds,  \eqref{EoM-H-sigma-sigma-dot} implies $\dot{\sigma}+3H\sigma \geq 0$. That is,  $\dot{\sigma}\geq 0$ at $\sigma=0$, where $\sigma$ changes sign. Therefore, as time evolves $\sigma$ can change sign from negative to positive values, while the reverse is not possible.


\section{Dipole Single Fluid Model}\label{sec:single-fluid}

To warm up for model building in dipole cosmology, in section \ref{sec:w-Lambda} we discuss the single fluid case. This is mainly a review of \cite{KMS, KMS-2}, with several further comments. In section \ref{sec:Dipole-radiation} we study the particular case where the fluid is radiation. The growth we identify in this case for negative $\beta$ will play a crucial role in  our later discussions of fluid mixture models, including  in  the dipole \lcdm\ model.

\subsection{Single Fluid: General Analysis and Remarks}\label{sec:w-Lambda}

One of the results of \cite{KMS, KMS-2} is that for dipole $w$-$\Lambda$ models, which involve a cosmological constant and a matter with constant EoS $w$, for $w>1/3$ the tilt $\beta$ can grow at late times.\footnote{See also \cite{Coley-1,Coley-2,Coley-3} for related observations in the context of various tilted cosmologies.} The relevant field equations are \cite{KMS, KMS-2}
\begin{equation}\label{w-Lambda-p-rho}
p=w\tilde{\rho}-\Lambda,\qquad \rho=\tilde\rho+\Lambda, \qquad -1< w\leq 1.
\end{equation}
For a generic $w$, \eqref{EoM-H-sigma} implies
\begin{subequations}\label{const-w-dipole-1}
\begin{align}
&H^2=\frac{\Lambda}{3}+\frac{A_0^2}{a^2}e^{-4b}+\frac{\tilde\rho}{3}\left[1+(w+1)\sinh^2\beta\right]+\frac19 \sigma^2\label{cosmic-acceleration-const-w}\\
&\dot{\beta}\big(\coth\beta-w \tanh \beta\big)=(3w-1)H-\frac23\sigma-\frac{2w A_0}{a(t)} e^{-2b}\tanh\beta \label{beta-growth-w-const}
\\
&\tilde\rho^{\frac{w}{1+w}} a e^{2b} \sinh\beta =  C = const.
\label{const-w-dipole-rho-X-beta}\\
&\sigma=\frac{C (1+w)}{2A_0} \tilde\rho^{\frac{1}{1+w}}\cosh\beta=\frac{(1+w)}{4A_0} a e^{2b}\ \tilde\rho\ \sinh2\beta \label{sigma-w-dipole}
\end{align}
\end{subequations}
where $C$ is an integration constant and in the above we assumed $w\neq -1$.

For non-negative cosmological constant $\Lambda\geq 0$, we make some observations below.
\begin{enumerate}
\item  As the universe expands, $a(t)$ grows while $\tilde\rho(t)$ and the shear $\sigma(t)$ go to zero. The universe isotropizes rapidly (exponentially fast for an accelerated expanding universe). This is consistent with the cosmic no-hair theorem \cite{Wald-cosmic-no-hair}, which still holds for  dipole cosmologies. 
\item Since $-1<w\leq 1$, for positive $\beta$ case $\coth\beta>1,$  $\tanh\beta <1$ and hence the coefficient of $\dot\beta$ term is always positive. While the last term in \eqref{beta-growth-w-const} does not have a definite sign, it becomes insignificant at late times due to the expansion. Therefore,  for $w>1/3$ the sign of  $\dot\beta$  is positive and the tilt can grow, see also  \cite{Coley-1, Coley-2}.
\item Cosmological constant $\Lambda$ does not explicitly appear in \eqref{beta-growth-w-const} and  $\beta$ growth depends mainly on  $w$ and the sign of $\beta$. The growth of $\beta$ can happen in accelerating/decelerating cosmologies. See \cite{KMS, KMS-2} for a detailed analysis.
\item As briefly outlined in the introduction, a key fact we will explore in this paper is that the sign of $\beta$ has crucial dynamical significance. In other words, $\beta \rightarrow -\beta$ is not a symmetry of the system, once sign of $A_0$ is fixed. This is because for a fixed value of $A_0$, the $z \leftrightarrow -z$ parity is broken, and therefore negative and positive values of $\beta$ are physically distinct. Under $\beta\to -\beta$ while holding $A_0$ fixed to a given positive\footnote{The discussion of the sign flip properties can be made with negative $A_0$ as well. The key point is that the system has a symmetry under $(A_0, \beta) \leftrightarrow -(A_0,\beta)$. But for concreteness, we work with positive $A_0$ throughout this paper.} number,  \eqref{cosmic-acceleration-const-w} does not change. But it is easy to see that  \eqref{const-w-dipole-rho-X-beta} and \eqref{sigma-w-dipole} together imply that the sign of $\sigma$ must change, and this means that the last two terms on the RHS of \eqref{beta-growth-w-const} have a sign change, while the rest of the terms do not. This shows that the full system of equations in dipole cosmology is {\it not} invariant under $\beta \rightarrow -\beta$.  Below we further explore this sign dependent dynamics. The crucial significance of the sign of $\beta$ in evolution does not seem to have been noted before.
\item In the previous literature \cite{Coley-1, Coley-2, Coley-3} and in \cite{KMS}, the main focus was on tilt growth and its asymptotic behavior at late times. However, in \cite{KMS} it was also noted that tilt growth may happen in the intermediate stages and in the early Universe. Depending on which of the three terms in the RHS of \eqref{beta-growth-w-const} is positive and dominant $\beta$ growth in different epochs may be driven by either of the three terms.

In the very early Universe and in the Big Bang era $a\to 0$. For a set of initial conditions which we suspect are fairly generic, $\sigma$ can be negative and can act as a dominant source for $\beta$ (see section \ref{sec:big-bang} for the specific  example of dipole \lcdm). For $\sigma<0$ we need to take $\beta<0$. From \eqref{cosmic-acceleration-const-w} we learn that $|\sigma|\leq 3H$. Numerically, we find that a large class of interesting initial condition have $\sigma\sim -3H$ in the early Universe. In such cases, starting from a small negative initial value for $\beta$ at the early times, last term in the RHS of \eqref{beta-growth-w-const} is negligible and  $(3w-1)H-2\sigma/3 \sim (3w+1)H$. 
So, we can get tilt growth for $w\geq -1/3$. However, $\sigma$ drops quite fast by the expansion (as $a^{-3}$ or faster) and the tilt growth stops when $\sigma$ has dropped to $3(3w-1)H/2$ value. These have been discussed in section \ref{sec:big-bang}.

If we start with negative $\beta$ and $w>0$, the last term in \eqref{beta-growth-w-const} can also contribute positively to tilt growth and can be dominant in some intermediate times. One may have power law or exponential expansions:
\begin{itemize}
    \item 
For a power law expansion $a\sim t^\alpha$, and $H\sim \alpha t^{-1}$ for $\alpha>1$ ($\alpha<1$) we have an accelerated (decelerated) expansion, and $\sigma\sim t^\gamma$ such that $|\sigma|<3H$ holds.  In intermediate times the main competition is between the $(3w-1)H$ and $w\tanh\beta e^{-2b}/a$ terms, 
$(3w-1)H-w\tanh\beta e^{-2b}/a\sim (3w-1)\alpha t^{-1}- w\tanh\beta t^{-\alpha}$. If $\alpha>1$ (accelerated expansion) the first term dominates at late times, while at intermediate times for $w>0, \beta<0$ or $w<0, \beta>0$ the $\tanh\beta$ term dominates. For decelerating $\alpha<1$ case, the second term dominates at late times and drives tilt growth if $w>0, \beta<0$ or $w<0, \beta>0$. Nonetheless, as we discussed in \cite{KMS} and briefly reviewed above, tilt growth driven by the $\tanh\beta$ term is very mild. 
\item For a (quasi)exponential expansion, $a\sim e^{H_0t}, H\sim H_0$ with a given (almost) constant $H_0$, $\sigma, \tanh\beta$ terms in \eqref{beta-growth-w-const} drop very fast at intermediate or late times and regardless of their sign, the dominant term is the $(3w-1)H$ term. Therefore,  in this case tilt growth in the asymptotic future  happens for $w\geq 1/3$.
\end{itemize}
All in all, tilt growth at some epoch is possible essentially for any $w$. As discussed, this growth may be driven by either of the three terms in the RHS of \eqref{beta-growth-w-const}.

\end{enumerate}

\subsection{Dipole \texorpdfstring{$\Lambda$}{}-radiation Model}\label{sec:Dipole-radiation}

\begin{figure}[H]
\centering
\subfloat[\label{fig:hnegative}]{\includegraphics[width = 7.5 cm]{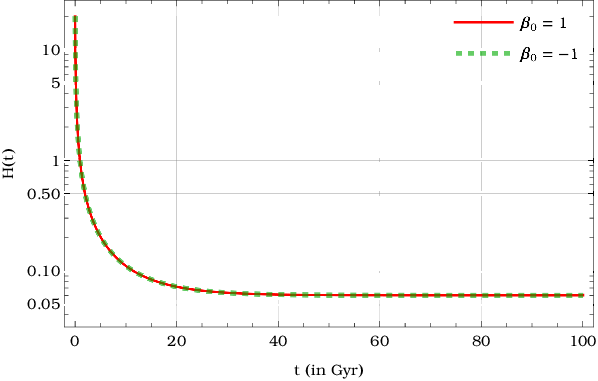}}\hfill
\subfloat[\label{fig:bnegative}]{\includegraphics[width = 7.5 cm]{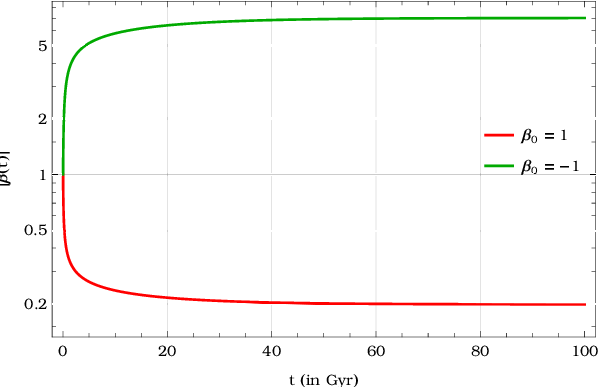}}\\
\subfloat[\label{fig:rhonegative}]{\includegraphics[width = 7.5 cm]{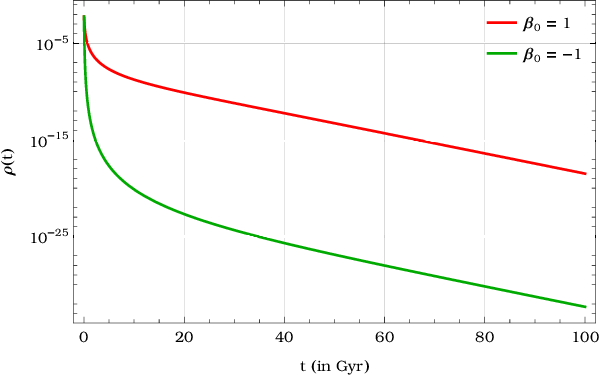}}\hfill
\subfloat[\label{fig:sigmanegative}]{\includegraphics[width = 7.5 cm]{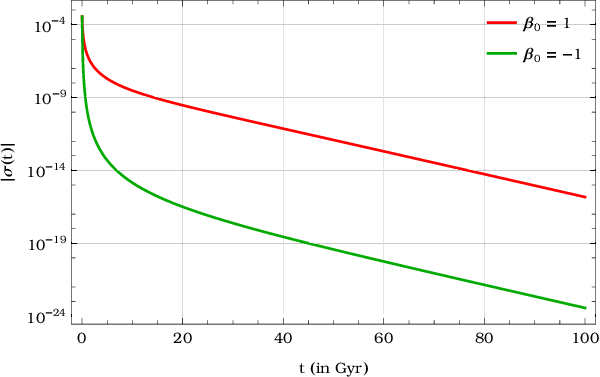}}\\
\caption{
Plots of Hubble parameter $H(t)$, absolute value of tilt $|\beta(t)|$, energy density $\rho(t)$ and shear $\sigma(t)$ for a fluid of EoS $w = 1/3$ with initial conditions $a_{0} = 0.05$, $b_{0} = 0$, $\rho_{0} = 0.006$. The cosmological constant has been chosen to be $\Lambda = 0.0109$. The green curves are for initial value $\beta_{0} =-1$ and red curves for $\beta_0=+1$. }
\label{fig:beta+-Lambda-rad}
\end{figure}

\begin{figure}[H]
    \centering
    \includegraphics[width = 7 cm, height = 5.5 cm]{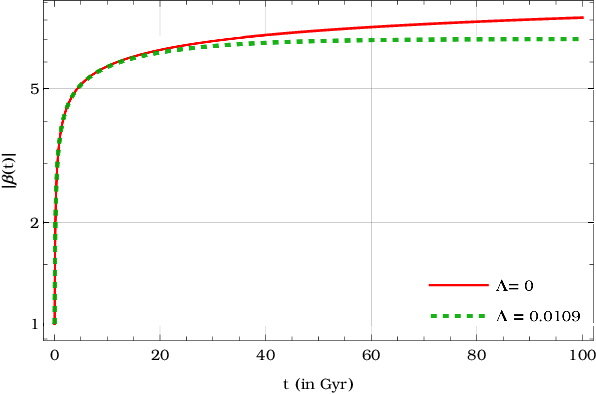}
    \caption{Comparing the tilt evolutions for universes with  and without  $\Lambda$ and for $\beta<0$.}
    \label{fig:beta-growth-Lambda-effect}
\end{figure}

As we observed, for positive $\beta$ the radiation $w=1/3$ case is at the borderline as far as asymptotic late-time tilt growth for $\beta>0$ is concerned. Given the above comments on the sign of $\beta$ and that radiation is an important component of any real cosmological model, we will explore this special case more closely. For $w=1/3$, 
\begin{subequations}\label{EOM-radiation-Lambda}
\begin{align}
&\tilde\rho= \frac{\rho_{r0}}{a^4\sinh^4\beta} e^{-8b}, \label{rho-radiation-beta}\\
&\sigma=\frac{2\rho_{r0}}{3A_0 a^{3}} \frac{\cosh\beta}{\sinh^3\beta}  e^{-6b}\label{sigma-radiation}\\
& H^2=\frac{\Lambda}{3}+ \frac{A_0^2}{a^2} e^{-4b} + \frac{\rho_{r0} e^{-8b}}{3a^4 \sinh^4\beta} (1+\frac43\sinh^2\beta)+ \frac{4\rho_{r0}^2 }{81 A_0^2}\frac{\cosh^2\beta e^{-12b}}{a^6\sinh^6\beta}.\label{H-radiation}\\
&\dot{\beta}\big(\coth\beta-\frac13 \tanh \beta\big)=-\frac{4\rho_{r0}}{9A_0 a^{3}}  \frac{\cosh\beta}{\sinh^3\beta}e^{-6b}-\frac{2A_0}{3a} \tanh\beta e^{-2b}\label{beta-growth-radiation}
\end{align}
\end{subequations}
where $\rho_{r0}$ is an integration constant. Given the above equations, some comments are in order.
\begin{enumerate}
    \item Eq.~\eqref{rho-radiation-beta} in which    $\rho_{r0}$ is an integration constant, exhibits the distinctive $a^{-4}$ radiation behavior.
\item Eq.~\eqref{sigma-radiation} shows the usual $a^{-3}$ falloff of the shear. Note also that sign of $\sigma$ and $\beta$ are the same; for positive (negative) $\beta$, $\sigma$ is positive (negative).
\item We have arranged the RHS of \eqref{H-radiation} in decreasing powers of the average scale factor $a$. As the universe expands the cosmological constant term dominates at very large $a$. The $a^{-2}$ term, the spatial curvature term, comes next. This is followed by the $a^{-4}$ term, the radiation contribution. Finally the $a^{-6}$ term has the distinctive form of the shear contribution. 
\item The possibility of tilt growth in $w=1/3$ case, is correlated with the sign of $\beta$, as is seen from \eqref{beta-growth-radiation}. We hence consider the positive and negative $\beta$ cases separately.
\begin{itemize}
    \item $\beta>0$. In this case $\coth\beta-\tanh\beta/3$ is strictly positive while the RHS of \eqref{beta-growth-radiation} is strictly negative. Therefore, $\dot\beta$ is always negative. Therefore, the tilt decreases in time. Due to expansion of the universe, the second term in the RHS rapidly dominates over the $a^{-3}$ term and this term governs $\beta$ dynamics. We have depicted the behavior of $\beta, \sigma, H$ for this case in red plots in Fig.~\ref{fig:beta+-Lambda-rad}.
    \item $\beta<0$. In this case $\coth\beta-\tanh\beta/3$ is strictly negative while both of the terms in the RHS of \eqref{beta-growth-radiation}, especially the dominant $\tanh\beta/a$ term,  is strictly positive. Therefore, $\dot\beta/\beta$ is always positive and the tilt can grow in time. The green plots in Fig.~\ref{fig:beta+-Lambda-rad} makes this explicit. 
\end{itemize}
\end{enumerate}  
    While $\beta$ grows for $\beta<0$ case and damps for $\beta>0$  as the plots clearly show, $\sigma, \rho$ and $H(t)$ are barely affected by the sign of $\beta$. Also, in Fig.~\ref{fig:beta-growth-Lambda-effect} we have studied effect of $\Lambda$  on the growth of the tilt for the $\beta<0$ case. As we see the absence of $\Lambda$  yields a slightly higher value of  the tilt at late times compared to the $\Lambda>0$ case. This may be understood recalling \eqref{beta-growth-radiation}, where the last term in the right-hand-side which goes as $1/a(t)$ is the dominant source for $\beta$, and that $a$ grows faster for $\Lambda>0$ case.\footnote{{This may be contrasted to the $w>1/3$ case discussed in \cite{KMS}, where the dominant source for $\beta$ growth is $(3w-1)H$ term. In this case bigger values of $H$ yields more $\beta$-growth. Therefore, positive $\Lambda$ case yields more growth than $\Lambda=0$ case.}}

To summarize this section, in the Dipole $\Lambda$-radiation cosmology, the tilt can grow if it starts with negative initial values. However, as the source of the tilt growth is exponentially falling off at late times we see a milder growth than the $w>1/3$ cases where the tilt is sourced by $(3w-1)H$ term (cf. \eqref{beta-growth-w-const}).


\section{Model Building: Dipole \texorpdfstring{$\Lambda$CDM}{LCDM} Model}\label{sec:Dipole-LCDM-basic}

In the usual FLRW setups, a cosmological model is generally defined by specifying different components of cosmic fluid and each component is typically defined by a constant equation of state (EoS), $p_i=w_i \rho_i$ with constant $w_i$. For example the standard model of cosmology at the background level, the so-called flat $\Lambda$CDM model, is specified by three sectors (1) dark energy which is formulated through a positive cosmological constant, a cosmic fluid with $w=-1$; (2) pressulreless matter with $w=0$ (consisting of dark matter and baryonic matter); (3) radiation (or relativistic species) with $w=1/3$. 

In \cite{KMS} however, it was argued that if we allow only for a single tilt parameter then the models one can build within the dipole cosmology framework are either limited to have a single cosmic fluid besides the cosmological constant or one should consider a generic time dependent EoS $w=w(t)$. The former option is very tight and does not allow for realistic cases where we want to allow for matter and radiation (and perhaps more). The latter option, while useful for making general statements, is quite unappealing as a model building setup. This is because a time-dependent equation of state means that we are specifying a {\it function} and not just a few numbers, and therefore the predictivity of this approach is quite limited.

In this paper, we will bypass this problem in a physically relevant and interesting way: There is no symmetry reason why different components of cosmic fluid should have the same tilt parameter. {Multiple fluids with different tilts seems to have been largely overlooked in the literature on tilted Bianchi models. Having multiple fluids and tilts is consistent with the symmetries of dipole cosmology as long as all the flows are along the same spatial direction (which we denote by $z$ in our notation). In an Appendix, we present explicit form of the stress tensor for the two-fluid case.} For example, radiation and matter sectors can have different tilt parameters, respectively $\beta_r, \beta_m$. We will call this, the {\it dipole $\Lambda$CDM model}. In principle one can let dark matter and baryonic matter sectors to also have different tilt factors $\beta_{\text{\tiny{DM}}}, \beta_{\text{\tiny{B}}}$. We will not show the analysis of the latter case here,  but will comment on it in the last section. 

Dipole $\Lambda$CDM model is described by the metric \eqref{DipoleMetric} and the energy momentum tensor,
\begin{equation}\label{dipole-LCDM-Tmunu}
\begin{split}
T^\mu{}_\nu = \text{diag}(-\Lambda-\rho_m-\rho_r, \Lambda +\frac{\rho_r}{3}, \Lambda & +\frac{\rho_r}{3}, \Lambda+\frac{\rho_r}{3}) + \left(\begin{array}{cccc}
 -{\cal T} &  ae^{2b} {\cal U} & 0 & 0 \\
-{\cal U}/(ae^{2b}) & {\cal T} & 0 & 0 \\
 0 & 0 & 0 & 0 \\
 0 & 0 & 0 & 0\\
\end{array}
\right)\\ 
{\cal T}:=\rho_m\sinh^2\beta_m+\frac43\rho_r\sinh^2\beta_r, & \qquad   {\cal U}:=\frac12(\rho_m\sinh2\beta_m +\frac43\rho_r\sinh2\beta_r)
\end{split}
\end{equation}
Thus the model involves 6 dynamical parameters $a, b; \rho_m, \beta_m; \rho_r, \beta_r$. The key assumption in model building using non-interacting fluid mixtures (both in FLRW as well as in dipole cosmology) is that the separate component fluids satisfy their own conservation equations. The time evolution is therefore described by Einstein and continuity equations, see appendix \ref{Appendix:consistency} for more details: 
\begin{subequations}\label{dipole-LCDM-EoM}
\begin{align}
&H^2=  \frac{\Lambda}{3}+\frac{A_{0}^{2}}{a^{2}}e^{-4b} + \frac13\rho_m \cosh^{2}{\beta_{m}}+ \frac13\rho_{r}(1 +\frac43\sinh^{2}{\beta_{r}})+\frac{\sigma^2}{9} \label{H2-dipole-LCDM}\\
& 4A_0\sigma=ae^{2b}(\rho_m \sinh2\beta_m+\frac43\rho_r \sinh2\beta_r)
\label{Shear-dipole-LCDM}\\ 
&\dot{\rho}_{m} + \rho_{m}\Big(3H + \tanh{\beta_{m}}\dot{\beta}_{m}-\frac{2A_0\tanh{\beta_{m}}}{a(t)}e^{-2b}\Big) = 0 \label{rho-m-LCDM}\\
&\dot{\rho}_{r} + 4\rho_{r}\Big(H + \frac{2}{3}\sigma + \coth{\beta_{r}}\dot{\beta}_{r}\Big) = 0 \label{p-r-dot-LCDM}\\
&\dot{\rho}_{r} + \frac43\rho_{r}\Big(3H + \tanh{\beta_{r}}\dot{\beta}_{r}-\frac{2A_0\tanh{\beta_{r}}}{a(t)}e^{-2b}\Big) = 0 \label{rho-r-LCDM}\\
& \coth{\beta_{m}}\dot{\beta}_{m} = -H - \frac{2}{3}\sigma  \label{p-m-dot-LCDM}.
\end{align}
\end{subequations}
The first two equations are coming from Einstein's equations and the last four are from continuity equations for radiation and pressureless matter. In other words, we have precisely enough equations to describe the six variables in the system.

\section{Dynamics of Dipole \texorpdfstring{$\Lambda$CDM}{LCDM} Cosmology}\label{sec:Dipole-LCDM-evolution}

In this section we explore the basic equations derived in the previous section. We first study the equations analytically and then in the next subsection we study the equations numerically and show the plots of some of the physical observables as a function of cosmic time $t$ and/or scale factor $a$.

\subsection{Analytical Treatment}\label{sec:analytic}

First, we note that for positive $\Lambda$, every term in the RHS of \eqref{H2-dipole-LCDM} is positive definite. So, each term can take a maximum value for a given $H$. In particular, 
\begin{equation}\label{X-domination}
   -3H\leq \sigma \leq 3H, \qquad \Lambda\leq 3H^2,\qquad A_0 \leq H \ a e^{2b},\qquad \rho_m\cosh^2\beta_m\leq 3H^2.
\end{equation}
The above together with \eqref{p-m-dot-LCDM} and \eqref{Shear-dipole-LCDM} yield,
\begin{equation}
    -3H\leq \coth\beta_m\ \dot{\beta}_m \leq H, \qquad -12H^2\leq {\rho}_m\sinh 2\beta_m+ {\rho}_r\sinh 2\beta_r \leq 12H^2
\end{equation}

One can integrate the last two equations \eqref{p-m-dot-LCDM} and \eqref{p-r-dot-LCDM} to obtain
\begin{subequations}\label{dipole-EoM-integrals}
\begin{align}
&a e^{2b} \sinh\beta_m=C_m  \label{beta-m-Cm}\\ 
&\rho_r= \frac{\rho_{r0}}{(a e^{2b}\sinh\beta_r)^{4} } \label{rho-r-rho0r}
\end{align}
\end{subequations}
where $C_m, \rho_{r0}$ are two integration constants. While $C_m$ can take positive or negative signs (depending on the initial sign of $\beta_m$), $\rho_{r0}$ should always be positive. Eq.\eqref{beta-m-Cm} shows that regardless of the sign, $|e^{2b}\sinh\beta_m|$ drops with the expansion of the universe as $1/a(t)$. Eq.\eqref{rho-r-rho0r} exhibits the usual fall off of the radiation energy density with the fourth power of $a e^{2b}\sinh\beta_r$ which may be viewed as the ``effective scale factor'' for radiation. We note that if $\sinh\beta_r$ drops with some power of scale factor $a(t)$, $\rho_r$ can have a faster than $a^{-4}$ fall off in the dipole cosmology.

From \eqref{dipole-LCDM-EoM} and \eqref{dipole-EoM-integrals} we learn
\begin{subequations}
\begin{align}
\frac{d}{dt} \ln\left({\rho_m}{a^3\cosh\beta_m}\right) &=\frac{2A_0}{C_m}\frac{\sinh^2\beta_m }{\cosh\beta_m}\label{rho-m-dynamics}\\
 2A_0\sigma&=C_m\rho_m \cosh\beta_m+\frac{4\rho_{r0}}{3} \frac{\cosh\beta_r}{({a e^{2b}\sinh\beta_r})^3}
\label{sigma-rho-m-rho-r}\\ 
 -(3\coth\beta_r-\tanh\beta_r)\dot\beta_r = \frac{d}{dt} \ln &\left(\frac{\cosh\beta_r}{\sinh^3\beta_r}\right) = 2\sigma+\frac{2\tanh\beta_r}{a} e^{-2b} \label{beta-r-LCDM}   
\end{align}\end{subequations}
Recalling \eqref{beta-m-Cm}, the RHS of \eqref{rho-m-dynamics} drops off at late times and hence $\rho_m\sim a^{-3}, \beta_m\sim 0$ at late times. The two terms on the RHS of \eqref{sigma-rho-m-rho-r} can have different late time fall off behavior and their sign is the same as the sign of respective tilts, first term is positive (negative) for positive (negative) $\beta_m$, and similarly for the second term. Therefore, if both $\beta_m, \beta_r$ are positive (negative) $\sigma$ is positive (negative), whereas if $\beta_m,\beta_r$ have different signs, depending on the initial conditions $\sigma$ can be positive or negative and it can change sign in time. As discussed, at late times the first term in \eqref{sigma-rho-m-rho-r} is expected to falloff like $a^{-3}$ and the second term may have faster or slower falloff, depending on behavior of $\beta_r$. If $\beta_r$ saturates a non-zero value at late times, this term will also drop like $a^{-3}$ and in general $\sigma\sim a^{-3}$, as in the un-tilted cases.  

Equation for $\beta_r$  \eqref{beta-r-LCDM} has exactly the same form as the radiation-$\Lambda$ dipole theory, with the difference that now $\sigma$ is also affected by the matter sector, as seen from \eqref{sigma-rho-m-rho-r}.
Note that while $\beta_r$ and $\beta_m$ can take either signs, their sign does not change in the course of evolution. Note also that $\sigma$ appears in the RHS of \eqref{beta-r-LCDM} and  acts as a source for $\beta_r$. Nonetheless, as discussed,  $\sigma$ term drops  as $a^{-3}$ or faster at late times and hence the dominant term which determines the late time evolution of $\beta_r$ is the $1/a(t)$ term. Therefore, recalling our discussions in the previous section and \eqref{beta-r-LCDM}, $\beta_r$ may grow if $\beta_r<0$. 

For completeness let us discuss different signs of $\beta$'s separately:
\begin{enumerate}
    \item $\beta_{r}, \beta_{m}>0$. In this case \eqref{sigma-rho-m-rho-r} shows $\sigma>0$ and \eqref{beta-r-LCDM} implies $\beta_r$ is a decreasing function. Therefore, the shear and both of the tilts fall off quickly with the expansion of the universe and in a relatively short time the Universe evolves to a $\Lambda$CDM model in an effectively FLRW setting. 
 \item $\beta_{r} > 0$, $\beta_{m} < 0$. In this case, the $C_m$ term in the RHS of \eqref{sigma-rho-m-rho-r} is negative while the second term is positive. So, depending on which term is bigger, $\sigma$ can be positive or negative. Given that the $C_m$ term drops like  essentially $a^{-3}$ and the radiation term drops faster, at late times $\sigma\to 0^-$ while both $\beta_m, \beta_r$ are decreasing in time. Like the previous case, the Universe evolves to usual $\Lambda$CDM model within FLRW setting. 
 \item  $\beta_{r} < 0$, $\beta_{m} > 0$. With the same argument as the previous case, at late times {  $\sigma\to 0^+$} while $|\beta_r|$ is increasing and $\beta_m$ is decreasing.
 \item $\beta_{r} , \beta_{m} < 0$. In this case \eqref{sigma-rho-m-rho-r} implies $\sigma<0$ and \eqref{beta-r-LCDM} implies $\dot\beta_r<0$ and hence $|\beta_r|$ grows in time while the shear quickly goes to zero. 
\end{enumerate} 
In summary, when $\beta_r<0$, $|\beta_r|$ grows at late times and when $\beta_m<0$ ($\beta_m>0$), $\sigma\to 0^-$ ($\sigma\to 0^+$), and in any case $|\beta_m|\to 0$. In the next subsection we will illustrate the above expectations through explicit numerical evolution and plots. 


\subsection{Numerical Treatment and Discussions}\label{sec:numeric}

As discussed above, when $\beta_r>0$ the system quickly evolves to a usual $\Lambda$CDM cosmology. Since we are interested in cases where we see a tilt growth, we therefore focus further on the $\beta_r<0$ cases. In this section we explore evolution of the system numerically. To make the analysis clearer, we may use the modified density parameters defined in \eqref{Omega's}, $\tilde\Omega_m, \tilde\Omega_r; \tilde\Omega_k, \tilde\Omega_\sigma, \Omega_\Lambda$, explicitly,
\begin{equation}\label{Omega's-LCDM}
   \begin{split}
 {\Omega}_{\Lambda}=\frac{\Lambda}{3H^2},\qquad H^2\tilde{\Omega}_k:=\frac{A_0^2}{a^2e^{4b}},\qquad \tilde{\Omega}_{\sigma}:=\frac{\sigma^2}{9H^2} \\
\tilde{\Omega}_{m}:=\frac{\rho_m}{3H^2}\cosh^2\beta_m, \qquad \tilde{\Omega}_{r}=\frac{\rho_r}{3H^2}\left(1+\frac43\sinh^2\beta_r\right),  
\end{split}\end{equation}
with the sum rule
\begin{equation}\label{Omega-LCDM-sum-rule}
\Omega_\Lambda+\tilde\Omega_m+ \tilde\Omega_r+\tilde\Omega_k+\tilde\Omega_\sigma=1, 
\end{equation} 
and that $\Omega$'s are  in $[0,1]$ range; see appendix \ref{Appendix:consistency} for definitions and more equations. So, if in a given epoch one of them approaches 1 and the rest are small, the Universe is dominated by that component in that epoch. In what follows we consider two cases, $\beta_r<0, \beta_m>0$ case and $\beta_r<0, \beta_m<0$ cases. In the former case shear $\sigma$ can change sign whereas in the latter $\sigma$ always remains negative. In both cases, as discussed, we see relative tilt growth ($\beta_r$ becoming more negative while $\beta_m$ goes to zero at late times). 

\begin{figure}[H]
\centering
\subfloat[\label{fig:betaLCDML}]{\includegraphics[width = 8 cm]{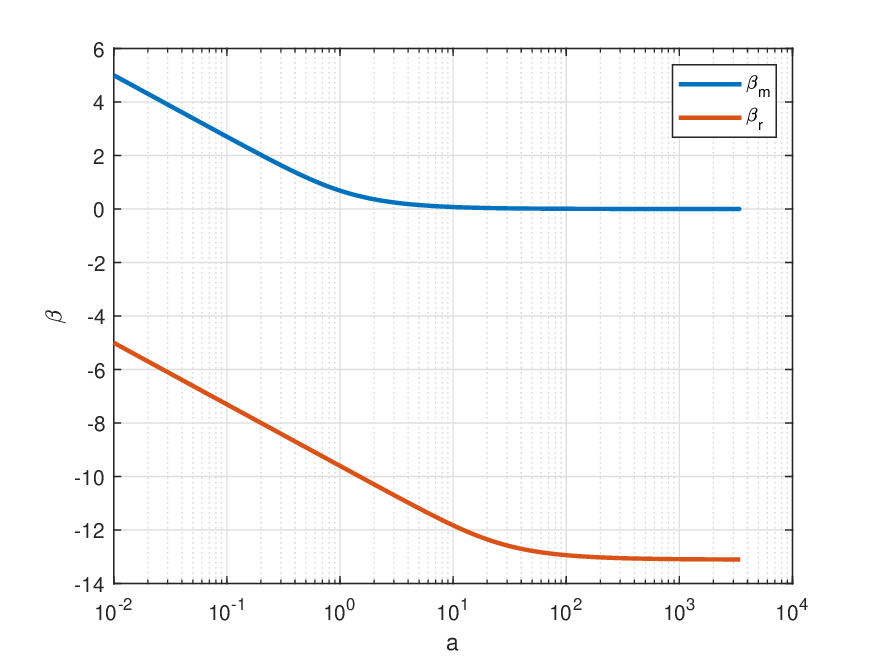}}\hfill
\subfloat[\label{fig:hLCDML}]{\includegraphics[width = 8 cm]{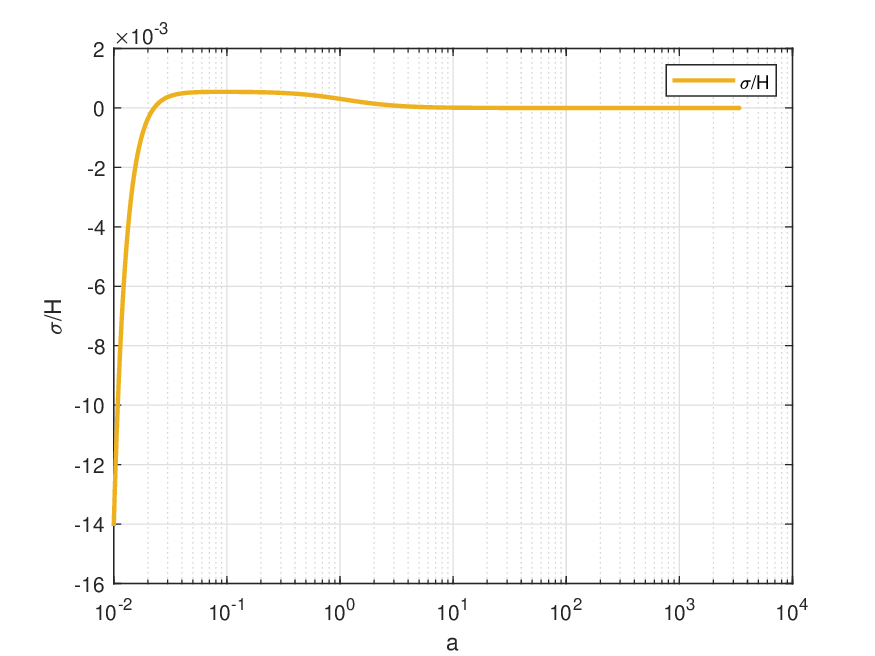}}\\
\subfloat[\label{fig:shearLCDML}]{\includegraphics[width = 8 cm]{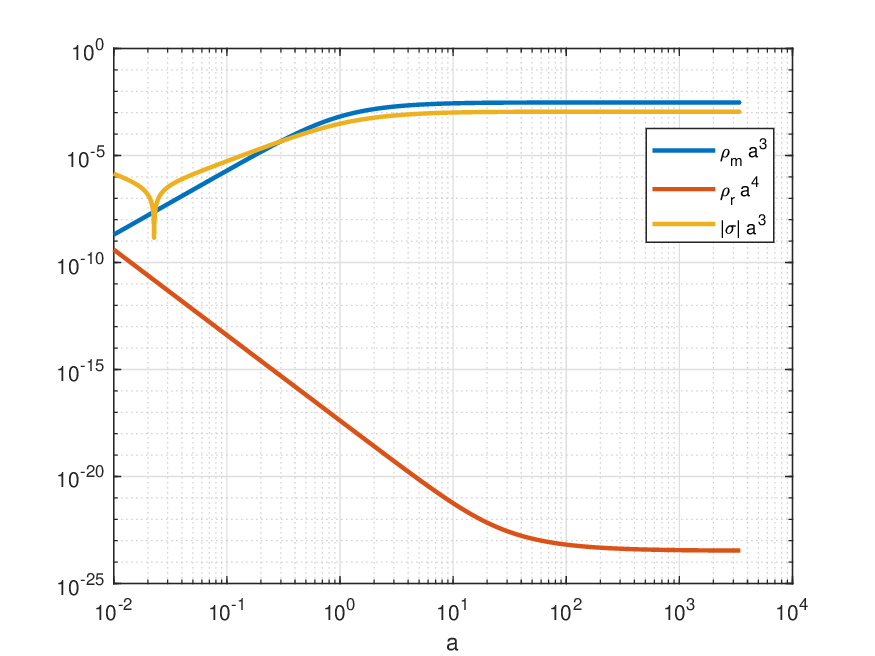}}\hfill
\subfloat[\label{fig:densityLCDML}]{\includegraphics[width = 8 cm]{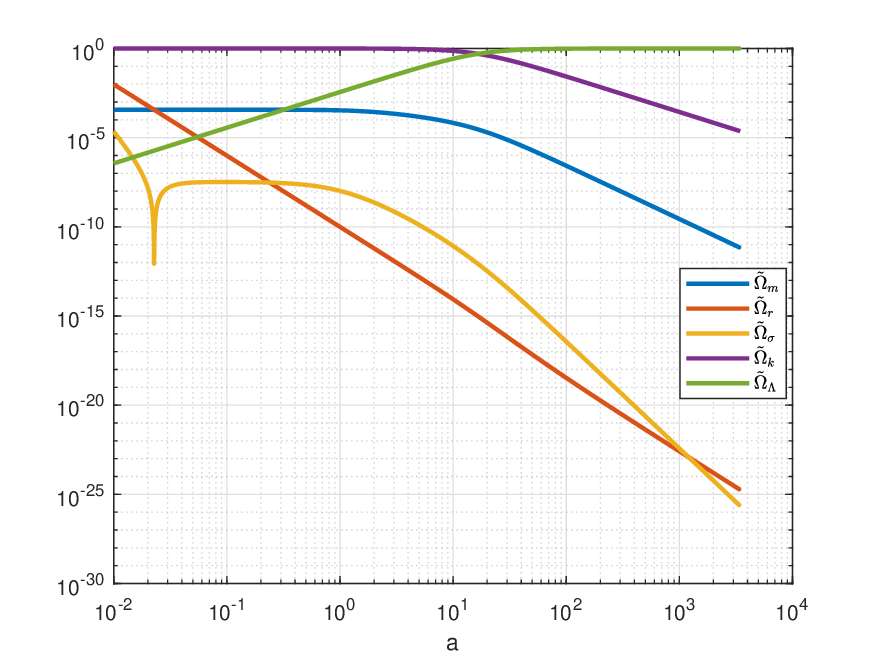}}\\
\caption{
Plots of tilts $\beta_r,\beta_m$, dimensionless ratio $\sigma/H$, scaled energy densities $\rho_r a^4, \rho_m a^3, |\sigma| a^3$, and the relative energy densities $\tilde\Omega_i$'s for $\beta_m>0, \beta_r<0$ case. These plots are for initial conditions  $b_{i} = 0$, $\rho_{m_i} = 0.002$, $\rho_{r_i} = 0.04$, $\beta_{m_i} = 5$, $\beta_{r_i} = -5$ and  $a_i=0.01$. The value of the cosmological constant is taken to be $\Lambda = 0.0109$.  }
\label{fig:beta+-Lambda-rad-Matter}
\end{figure}

\begin{figure}[H]
\centering
\subfloat[\label{fig:shearLowL}]{\includegraphics[width = 7.5 cm]{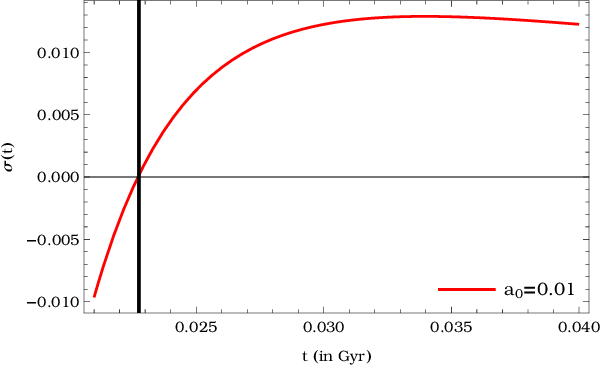}}\hfill
\subfloat[\label{fig:shearHighL}]{\includegraphics[width = 7.5 cm]{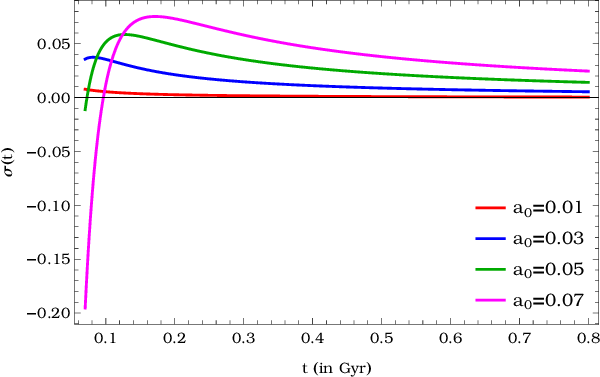}}
\caption{
When $\beta_m>0, \beta_r<0$, the shear can change sign at some early time.  }
\label{fig:beta+-Lambda-rad-MatterShear}
\end{figure}

\begin{figure}[H]
\centering
\subfloat[\label{fig:betaMMLCDML}]{\includegraphics[width = 8 cm]{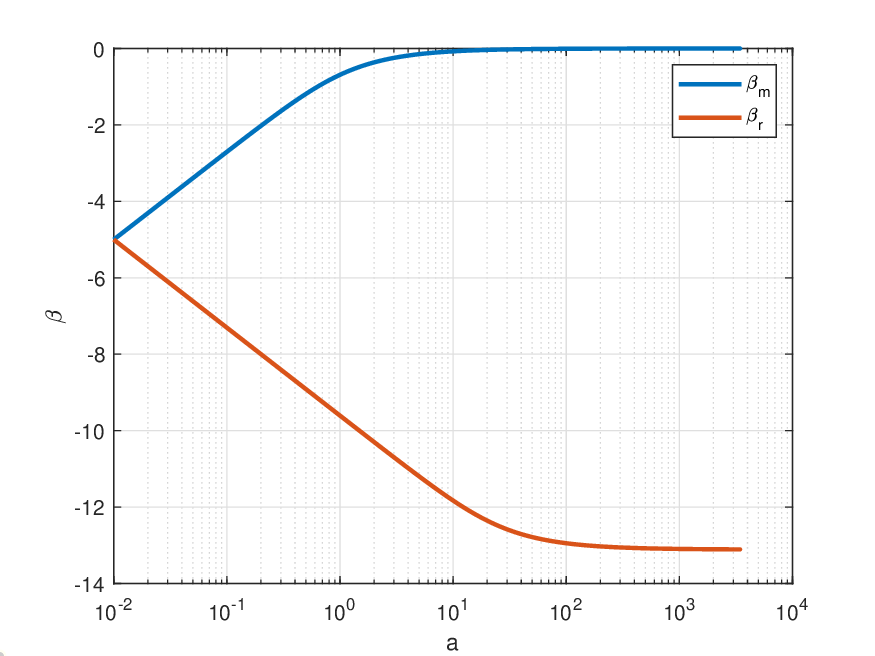}}\hfill
\subfloat[\label{fig:hMMLCDML}]{\includegraphics[width = 8 cm]{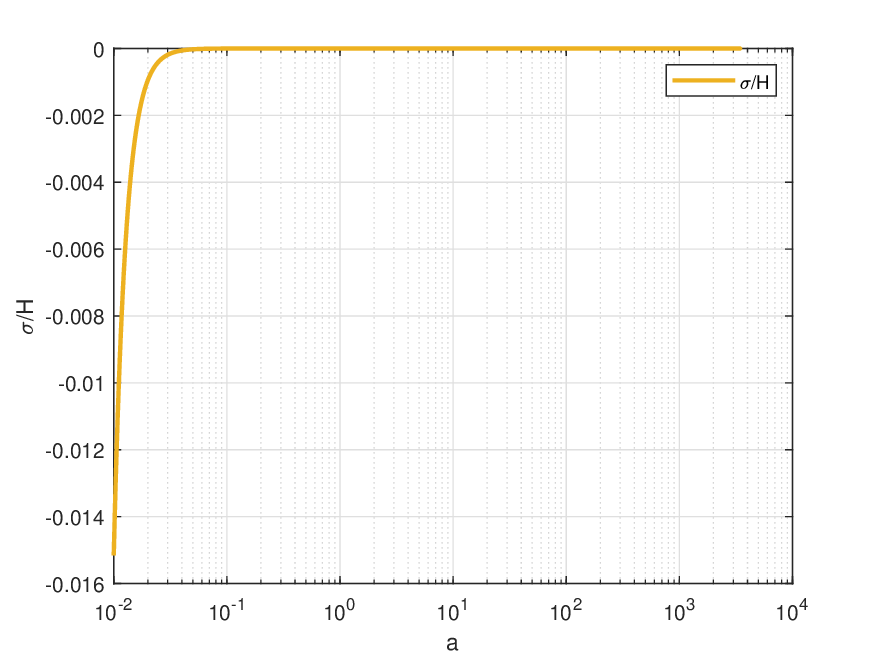}}\\
\subfloat[\label{fig:densityMMLCDML}]{\includegraphics[width = 8 cm]{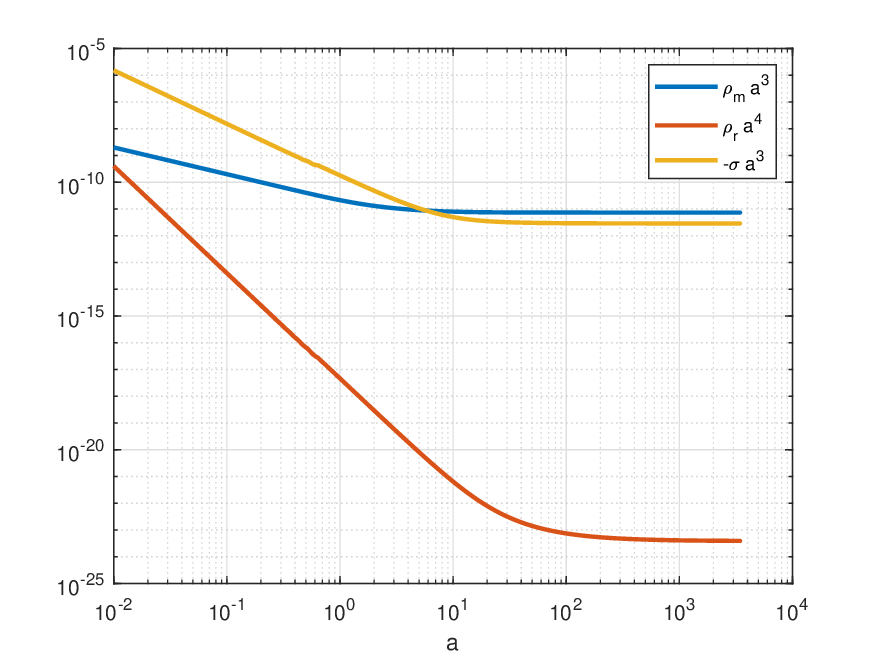}}\hfill
\subfloat[\label{fig:shearMMLCDML}]{\includegraphics[width = 8 cm]{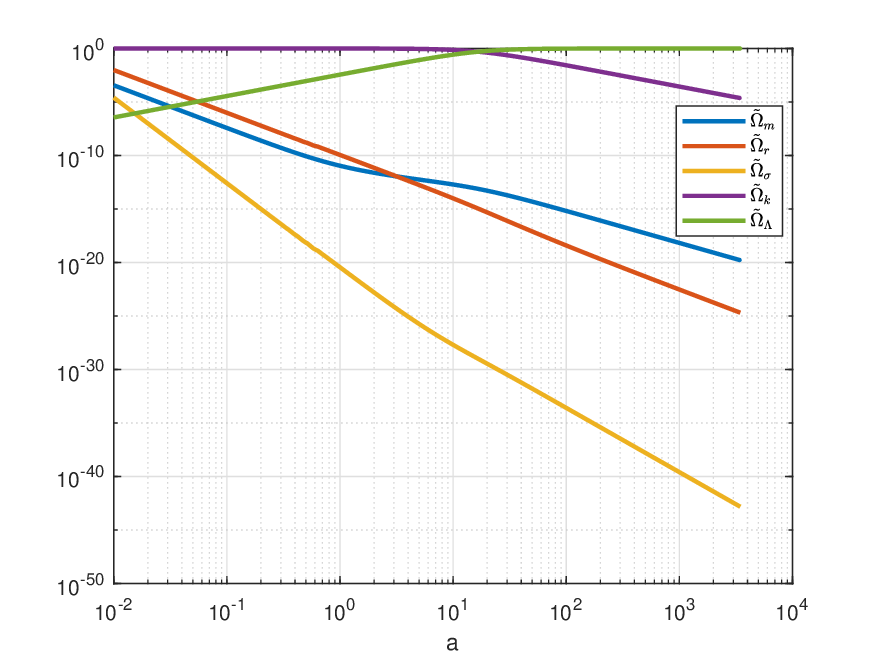}}\\
\caption{
Plots of  $\beta_r, \beta_m$, $\sigma/H$, scaled energy densities and relative energy densities for $\Lambda$CDM model for $\beta_m<0, \beta_r<0$ case governed by \eqref{dipole-LCDM-EoM}  with initial conditions  $b_{i} = 0$, $a_i=0.01$, $\rho_{m_i} = 0.002$, $\rho_{r_i} = 0.04$, $\beta_{m_i} = -5$, $\beta_{r_i} = -5$. The value of the cosmological constant is taken to be $\Lambda = 0.0109$. }
\label{fig:beta--Lambda-rad-Matter}
\end{figure}

\paragraph{$\beta_r<0, \beta_m> 0$ case.} As discussed, besides growth in $|\beta_r|$ and fall off of $\beta_m$ and $\sigma$, we expect $\sigma$ to change sign (from negative to positive) at some point in time. These features are depicted in Fig.~\ref{fig:beta+-Lambda-rad-Matter} and Fig.~\ref{fig:beta+-Lambda-rad-MatterShear}. Some comments are in order:
\begin{itemize}
    \item Plot Fig.~\ref{fig:betaLCDML} shows that $\beta_m$ goes to zero while $\beta_r$ can have a mild growth asymptoting to a non-zero negative value, in accord with our analytical discussions. Note in particular that the saturation of $\beta_r$ is expected in an expanding Universe where shear decreases at late times.
    \item Plot Fig.~\ref{fig:hLCDML} shows that $\sigma$ can change sign and that it has a maximum point where $\dot{\sigma}=0$. Recalling \eqref{EoM-H-sigma-sigma-dot}, we note that $\dot{\sigma}+3H\sigma\geq 0$. That is, when $\sigma=0$, $\dot{\sigma}\geq 0$, and when $\dot{\sigma}=0$, $\sigma\geq 0$.  Therefore, $\sigma$ can only change sign from negative to positive, and not the reverse. This feature is explored more closely in Fig.~\ref{fig:beta+-Lambda-rad-MatterShear}.
    \item Moreover, the RHS of \eqref{EoM-H-sigma-sigma-dot} vanishes asymptotically, as $\rho_r, \rho_m$ go to zero and for $\Lambda$-term, $\rho+p=0$. So, one expects $\sigma a^3$ to go to a constant asymptotically, as is seen in Fig.~\ref{fig:shearLCDML}.
    \item As Fig.\ref{fig:beta+-Lambda-rad-MatterShear} shows, while $\sigma\to 0^+$ at late times, the shear can change sign at some early time.  The sign change time can be tuned.  For a given set of initial conditions for other parameters as in Fig.~\ref{fig:beta+-Lambda-rad-Matter}, the sign change happens at a later time as we increase $a_{i}$.
    \item The cups in $|\sigma| a^3, \tilde\Omega_\sigma$ plots Figs.~\ref{fig:shearLCDML} and \ref{fig:densityLCDML} are due the fact that $\sigma$ changes sign, as depicted in Fig.~\ref{fig:hLCDML}. 
    \item As Fig.~\ref{fig:shearLCDML} shows $\rho_r a^4, \rho_m a^3, |\sigma| a^3$ approach to constant values in asymptotic future. This behavior may be analytically argued for noting \eqref{rho-r-rho0r}, \eqref{rho-m-dynamics}, \eqref{sigma-rho-m-rho-r}, respectively. In particular, from \eqref{sigma-rho-m-rho-r} one learns that asymptotically
\begin{equation}\label{sigma-rho-m-rho-r-asymptotic}
     2A_0\sigma a^3 \simeq C_m\rho_m a^3 +\frac{4\rho_{r0}}{3} \frac{\cosh\beta^\infty_r}{\sinh^3\beta^\infty_r}
\end{equation}
where $\beta^\infty_r$ is the asymptotic value of radiation tilt (cf. Fig.~\ref{fig:betaMMLCDML}).
\item Since $\beta_m<0$, for this case $C_m<0$ (cf. \eqref{beta-m-Cm}). And since $\beta_r^\infty\simeq -13$ for the values we consider in the plot, the last term in \eqref{sigma-rho-m-rho-r-asymptotic} is much smaller than the other two terms. Therefore, for our choice of initial values, asymptotically, $2A_0\sigma a^3 \simeq C_m\rho_m a^3$, while $\rho_m a^3$ is expected to be slightly larger than $|\sigma| a^3$.
    \item For the values of parameters chosen in these plots, curvature term is relatively large ($A_0^2/\Lambda\sim {\cal O}(1)$). This leads to the features seen in Fig.~\ref{fig:densityLCDML} that a ``curvature dominated'' Universe evolves to a $\Lambda$-dominated Universe. For chosen set of initial values we do not see a matter or radiation dominated Universe in the last few e-folds till now $(a=1$). 
    \end{itemize}

\paragraph{$\beta_m<0, \beta_r<0$ case.} The evolution for this case is depicted in Fig.~\ref{fig:beta--Lambda-rad-Matter}. Some comments are in order:
\begin{itemize}
    \item 
In this case, $|\beta_r|$ can grow (seemingly asymptoting to a constant value) while $|\beta_m|$ always falls off in time. 
\item The shear always remains negative, has no extremum point and tends to zero at late times.  
\item As Fig.~\ref{fig:densityMMLCDML} shows $\rho_r a^4, \rho_m a^3, \sigma a^3$ asymptote to constant values. Recalling \eqref{sigma-rho-m-rho-r-asymptotic} and discussions below that, in this case $|\sigma| a^3$ is expected to be slightly bigger than $\rho_m a^3$.
\item As in the previous case in Fig.~\ref{fig:beta+-Lambda-rad-Matter}. due to a unrealistically (compared to Planck values \cite{Planck-2018}) large $A_0$, $A_0^2/\Lambda\sim {\cal O}(1)$, a curvature dominated Universe evolves to a $\Lambda$-dominated Universe. For chosen set of initial values we do not see a matter or radiation dominated Universe in the last few e-folds till now $(a=1$). We will discuss more ``realistic'' (as in Planck-compatible) values for parameters in a later section. 
\end{itemize}

\subsection{Tilt Growth for \texorpdfstring{$\Lambda=0$}{}}\label{sec:dipole CDM}

\begin{figure}[H]
\centering
\subfloat[\label{fig:shearLow}]{\includegraphics[width = 8 cm]{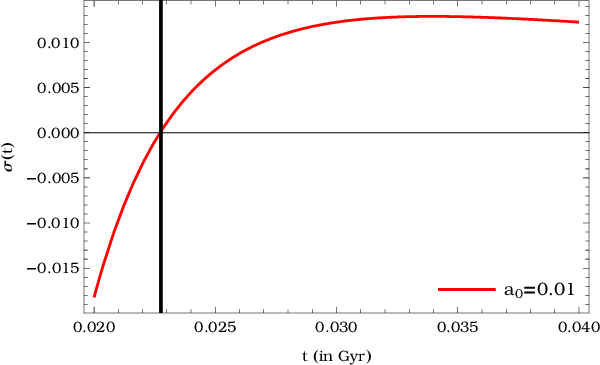}}\hfill
\subfloat[\label{fig:shearHigh}]{\includegraphics[width = 8 cm]{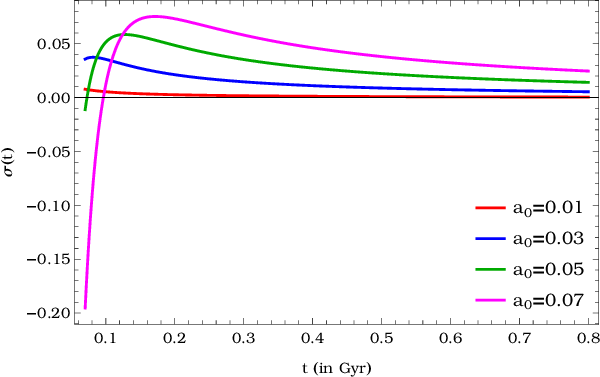}}
\caption{
For $\beta_m>0, \beta_r<0$ shear can change sign at early times for $\Lambda=0$ case. As we increase the initial value of the scale factor the change of sign happens at a larger time. At late times $\sigma\to 0^+$. }
\label{fig:beta+-rad-MatterShear}
\end{figure}

\begin{figure}[H]
\centering
\subfloat[\label{fig:hMMLCDM}]{\includegraphics[width = 8 cm]{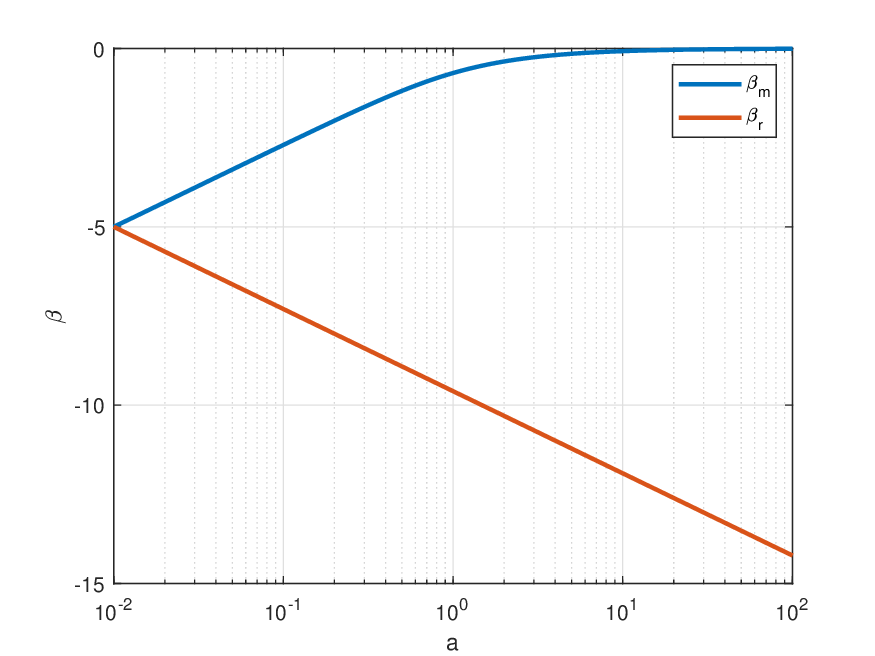}}\hfill
\subfloat[\label{fig:densityMMLCDM}]{\includegraphics[width = 8 cm]{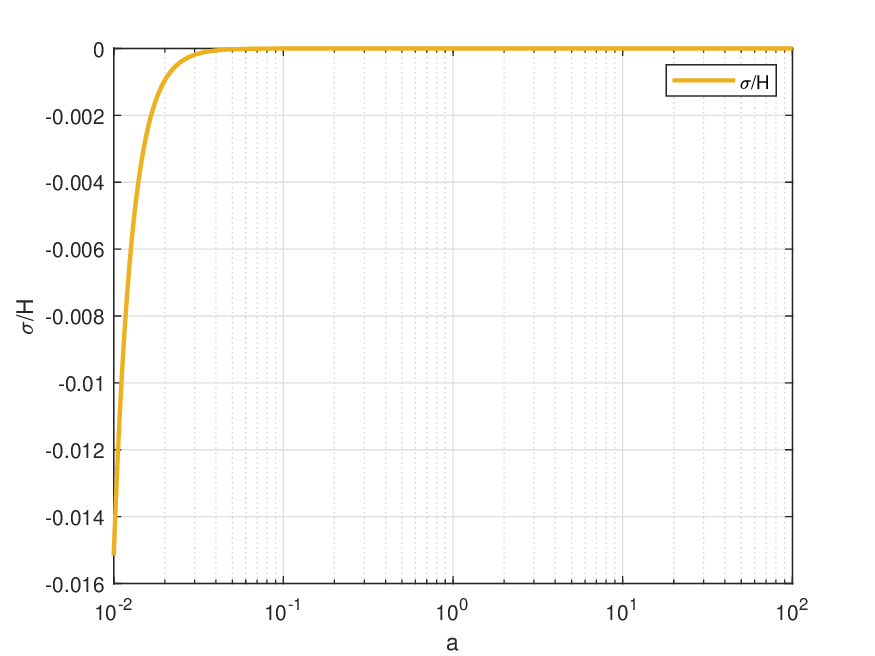}}\\
\subfloat[\label{fig:betaMMLCDM}]{\includegraphics[width = 8 cm]{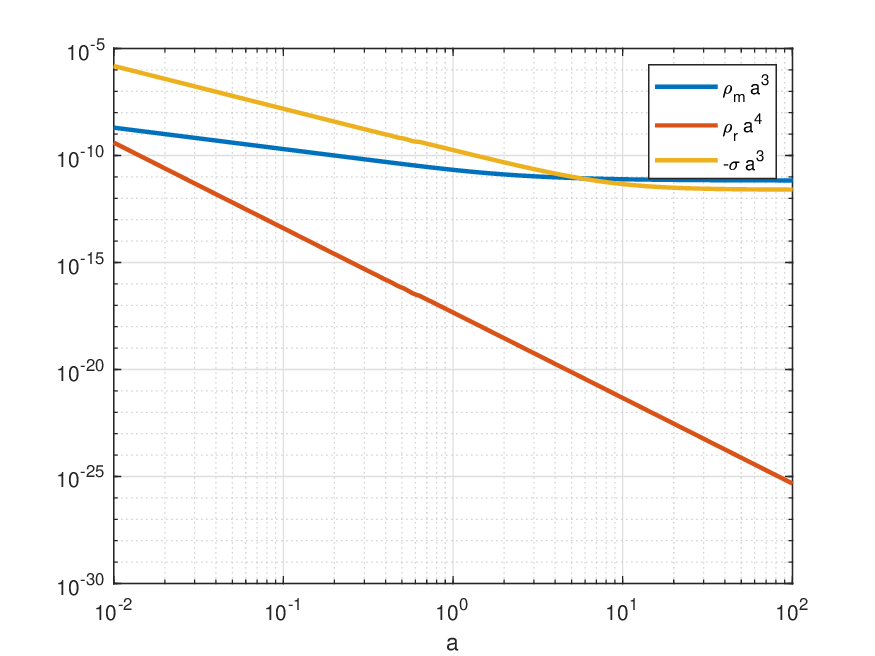}}\hfill
\subfloat[\label{fig:shearMMLCDM}]{\includegraphics[width = 8 cm]{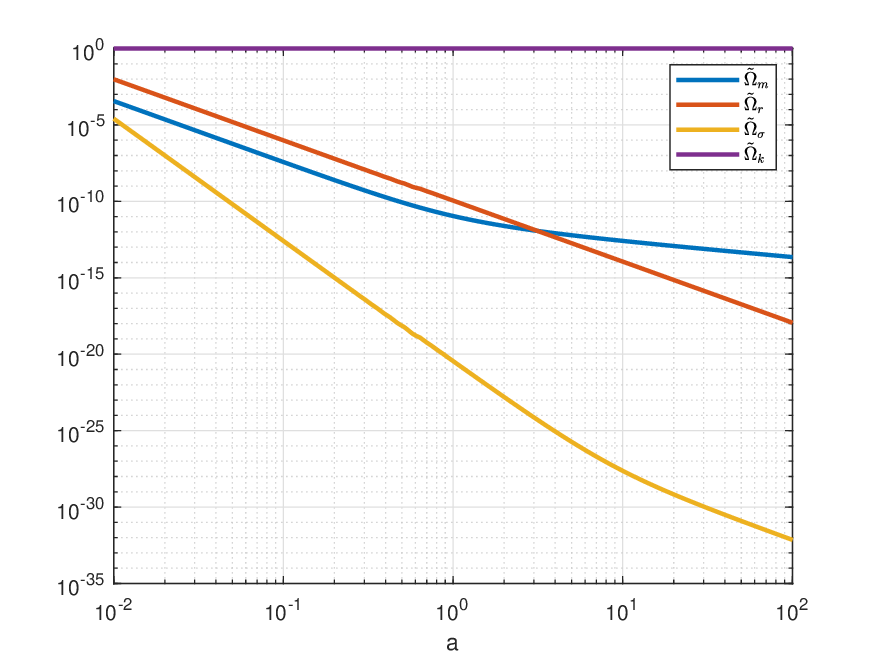}}\\
\caption{lots of tilts $\beta_r,\beta_m$, $\sigma/H$, scaled energy densities $\rho_r a^4, \rho_m a^3, |\sigma| a^3$ and relative energy densities $\tilde\Omega$'s for $\Lambda=0$ and $\beta_m<0, \beta_r<0$ case. Initial conditions  are taken as $b_{i} = 0$, $a_i=0.01$, $\rho_{m_i} = 0.002$, $\rho_{r_i} = 0.04$, $\beta_{m_i} = -5$, $\beta_{r_i} = -5$.}
\label{fig:beta--rad-Matter}
\end{figure}

The analysis in section \ref{sec:analytic} are quite generic and give the picture on how the tilts and shear behave as a function of scale factor $a(t)$. However, the late time behaviour of the scale factor, whether it is power law or exponential, depends on the presence and sign of $\Lambda$. In section \ref{sec:numeric} we focused on the $\Lambda>0$  case and in this part we present plots for $\Lambda=0$ case. Here, paralleling the  previous subsection, we again consider $\beta_r<0, \beta_m>0$ and $\beta_r, \beta_m <0$ cases. The main difference between the $\Lambda>0$ of previous subsection and the $\Lambda =0$ here is that at late times $a(t)\sim e^{H_0 t}, 3H_0^2=\Lambda$ for $\Lambda>0$ case, whereas for $\Lambda=0$, we have a power-law behaviour $a\sim t^{2/3}$.    
\begin{figure}[H]
\centering
\subfloat[\label{fig:hLCDM}]{\includegraphics[width = 8 cm]{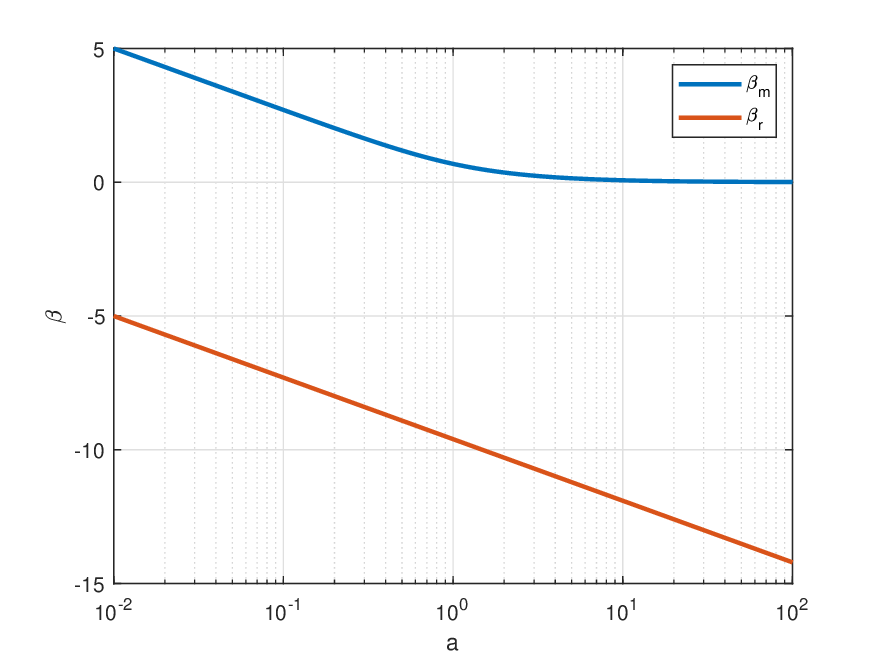}}\hfill
\subfloat[\label{fig:densityLCDM}]{\includegraphics[width = 8 cm]{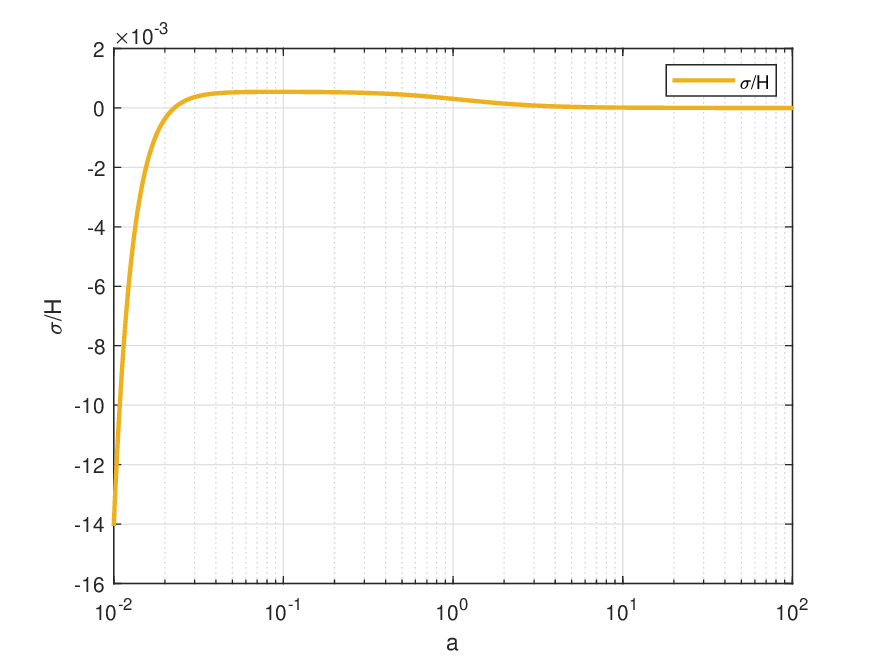}}\\
\subfloat[\label{fig:betaLCDM}]{\includegraphics[width = 8 cm]{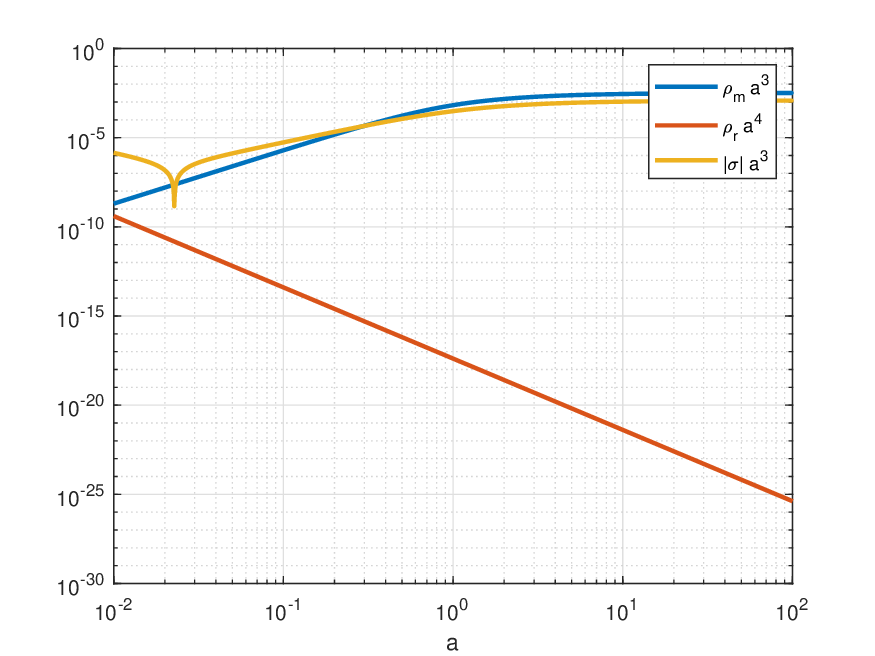}}\hfill
\subfloat[\label{fig:shearLCDM}]{\includegraphics[width = 8 cm]{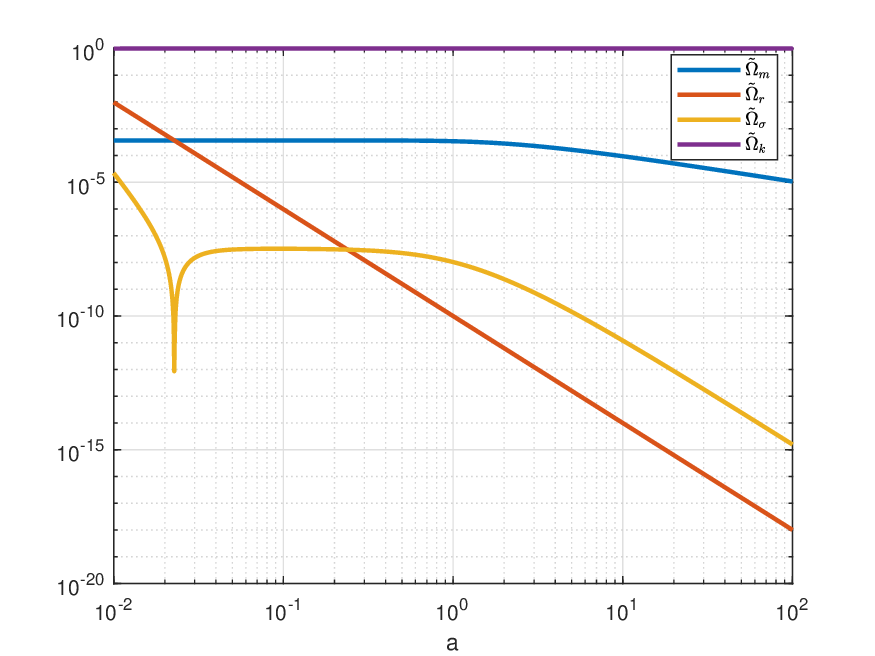}}\\
\caption{
Plots of tilts $\beta_r,\beta_m$, $\sigma/H$, scaled energy densities $\rho_r a^4, \rho_m a^3, |\sigma| a^3$ and relative energy densities $\tilde\Omega$'s for $\Lambda=0$ and $\beta_m>0, \beta_r<0$ case. Initial conditions  are taken as $b_{i} = 0$, $a_i=0.01$, $\rho_{m_i} = 0.002$, $\rho_{r_i} = 0.04$, $\beta_{m_i} = 5$, $\beta_{r_i} = -5$.} 
\label{fig:beta+-rad-Matter}
\end{figure}
\paragraph{$\beta_r<0, \beta_m>0$ case.} Comparing Fig.~\ref{fig:beta+-rad-Matter}  with Fig.~\ref{fig:beta+-Lambda-rad-Matter} we see that relative $\beta$ growth is essentially not affected by the presence/absence of $\Lambda$, while $\sigma, \rho$ drop faster for $\Lambda>0$ case, which is a result of exponential vs power-law growth of $a(t)$ in the two cases. As analytically expected, see \eqref{sigma-rho-m-rho-r-asymptotic}, $\rho_m a^3, \sigma a^3$ asymptote to constant values. However, $|\beta_r|$ and $\rho_r a^4$ continue their evolution, respectively to larger and smaller values. Moreover, as Fig.~\ref{fig:shearLCDM} shows and in the absence of $\Lambda$-term, for our set of initial values, we have ``curvature dominance" in the last few e-folds all the way to asymptotic future. Note that the cusp in $|\sigma| a^3$ and $\tilde\Omega_\sigma$ plots is due to the fact that $\sigma$ changes sign. To highlight this change of sign, in Fig.~\ref{fig:beta+-rad-MatterShear} we have explored more closely $\sigma$ as a function of cosmic time $t$. This figure may be compared to Fig.~\ref{fig:beta+-Lambda-rad-MatterShear}.
\paragraph{$\beta_r<0, \beta_m<0$ case.} We finally compare Fig.~\ref{fig:beta--rad-Matter}  with Fig.~\ref{fig:beta--Lambda-rad-Matter}. The notable features here are while $\rho_m a^3, \sigma a^3$ asymptote to constant values, $|\beta_r|$ and $\rho_r a^4$ continue their evolution, respectively to larger and smaller values. 

\subsection{Big Bang in Dipole Cosmology}\label{sec:big-bang}

\begin{figure}[H]
\centering
\subfloat[\label{fig:backbeta1}]{\includegraphics[width = 8 cm]{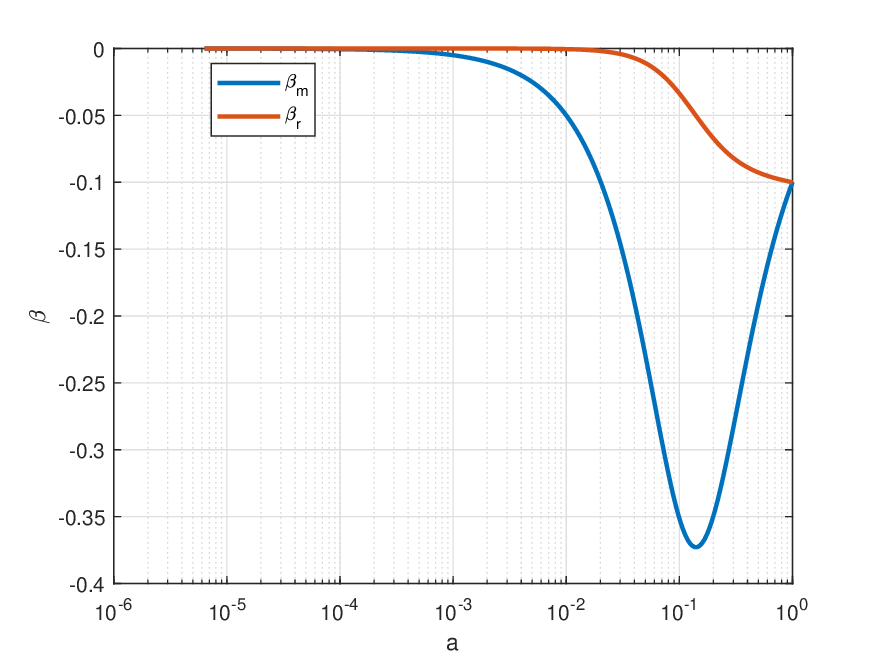}}\hfill
\subfloat[\label{fig:backsig1}]{\includegraphics[width = 8 cm]{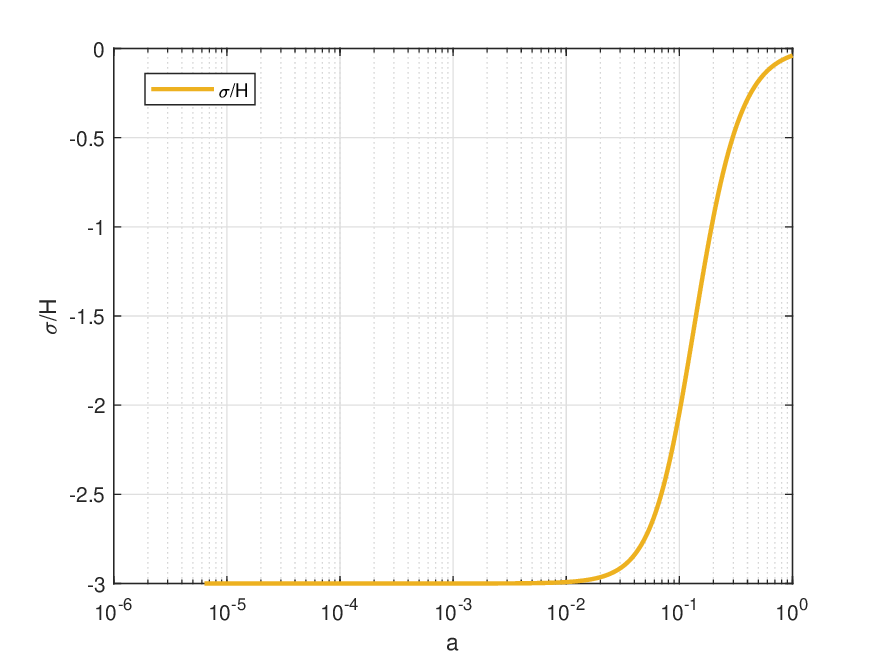}}\\
\subfloat[\label{fig:backa1}]{\includegraphics[width = 8 cm]{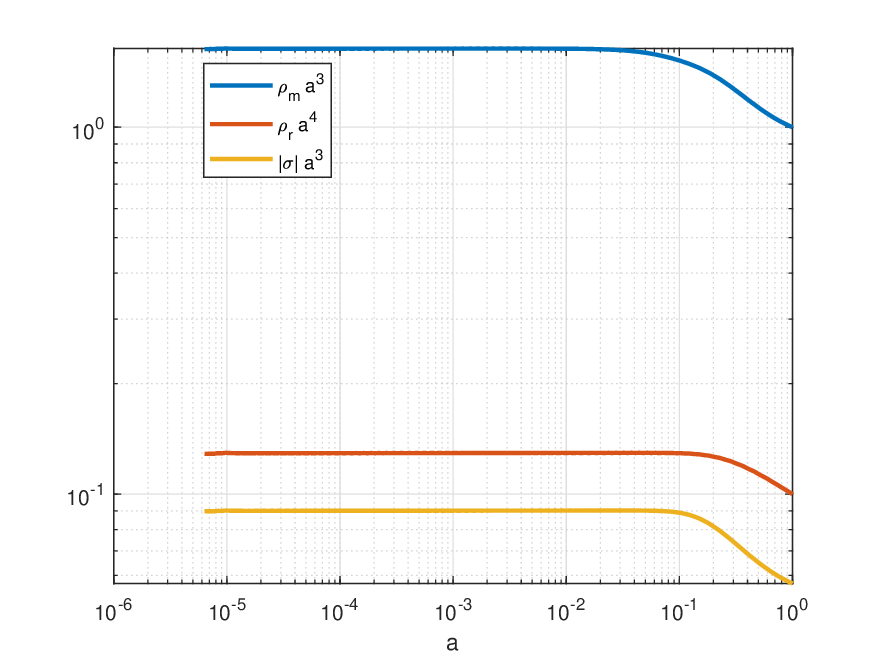}}\hfill
\subfloat[\label{fig:backOmega1}]{\includegraphics[width = 8 cm]{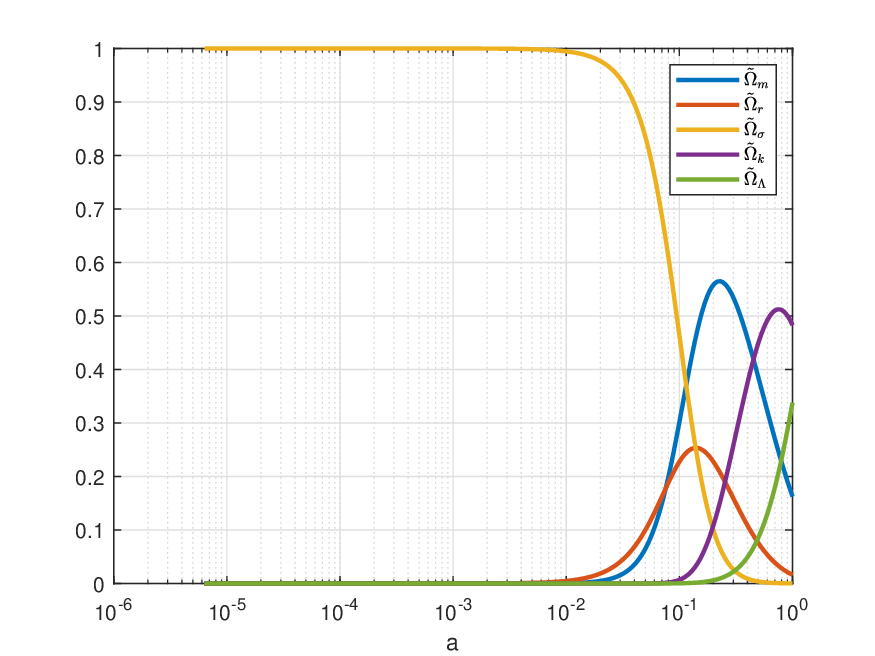}}\\
\caption{
Plots for evolution backward in time, from today $t=0$ to the Big Bang era  $a\to 0$ with  $\rho_m(0)=1,\rho_r(0)=0.1,\beta_m(0)=\beta_r(0)=-0.1$ at $a=1$. 
Fig.~\ref{fig:backbeta1} shows matter and radiation tilts vs. scale factor $a$. In Fig.~\ref{fig:backsig1} we have plotted  dimensionless ratio $\sigma/H$ vs $a$. It shows that how the shear which can be large at initial times goes to zero in the current epoch. Finally Fig.~\ref{fig:backOmega1} shows how respective ratios of matter, radiation and shear densities evolve as a function of the scale factor $a$. }
\label{fig:back1}
\end{figure}

\begin{figure}[H]
\centering
\subfloat[\label{fig:backbeta2}]{\includegraphics[width = 8 cm]{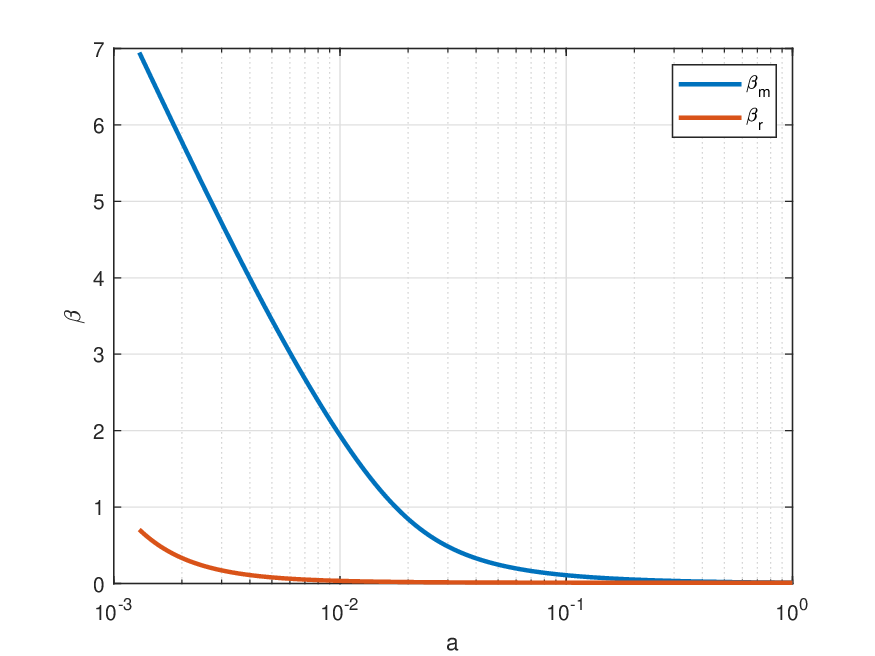}}\hfill
\subfloat[\label{fig:backsig2}]{\includegraphics[width = 8 cm]{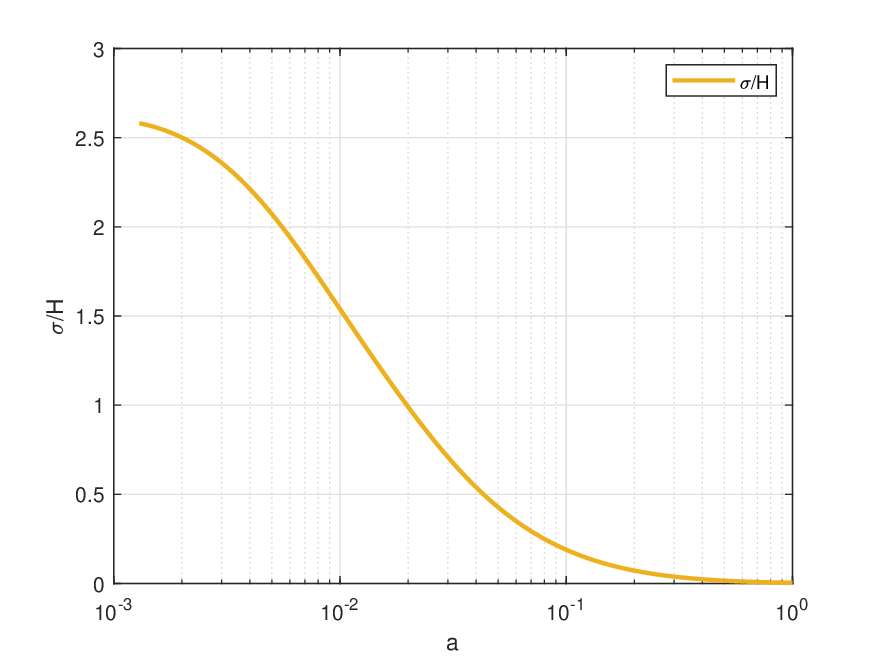}}\\
\subfloat[\label{fig:backrho2}]{\includegraphics[width = 8 cm]{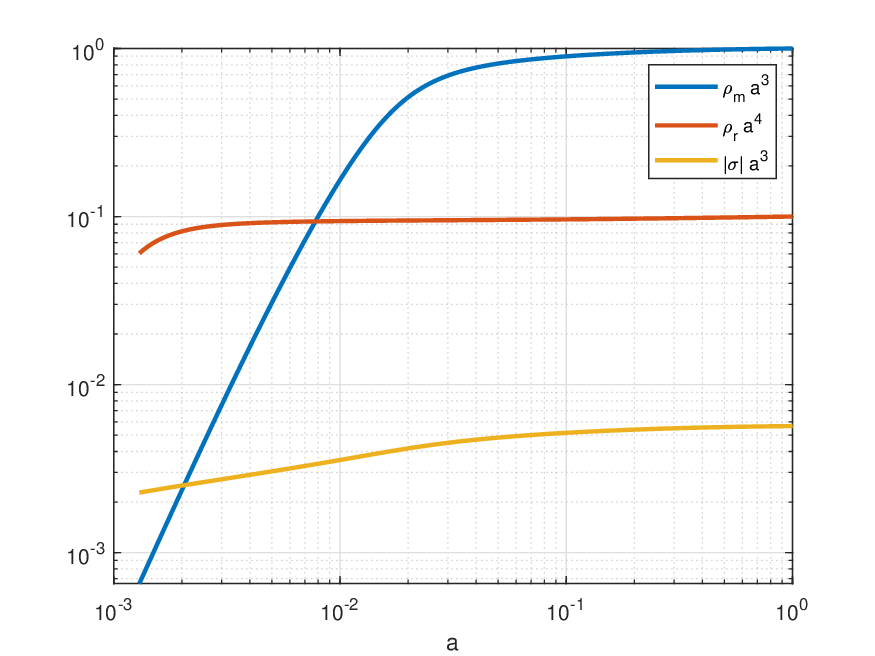}}\hfill
\subfloat[\label{fig:backOmega2}]{\includegraphics[width = 8 cm]{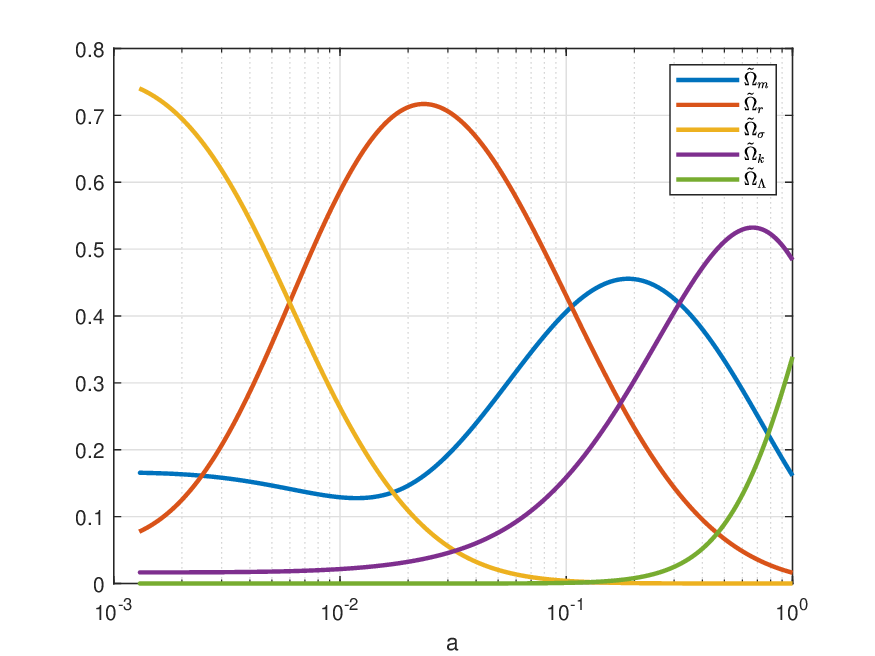}}
\caption{
Plots of the rapidity obtained with  $\rho_m(0)=1,\rho_r(0)=0.1,\beta_m(0)=0.01,\beta_r(0)=0.01$ at $a=1$ and solved backward in time; evolving from far in the past (near Big Bang) to current times. 
}\label{fig:back2}
\end{figure}

\begin{figure}[H]
\centering
\subfloat[\label{fig:backbetaLCDM}]{\includegraphics[width = 7.5 cm]{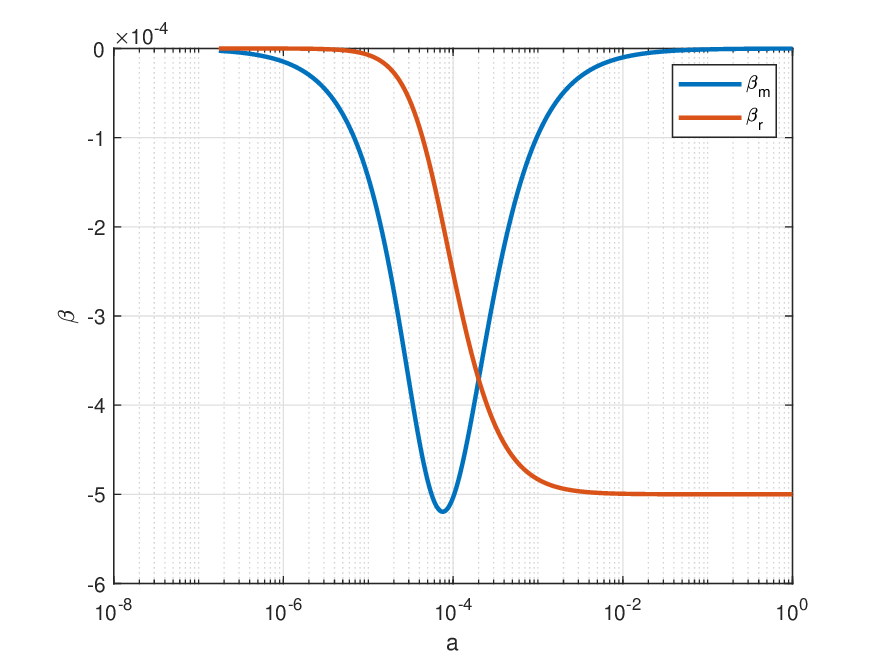}}\hfill
\subfloat[\label{fig:backsigLCDM}]{\includegraphics[width = 7.5 cm]{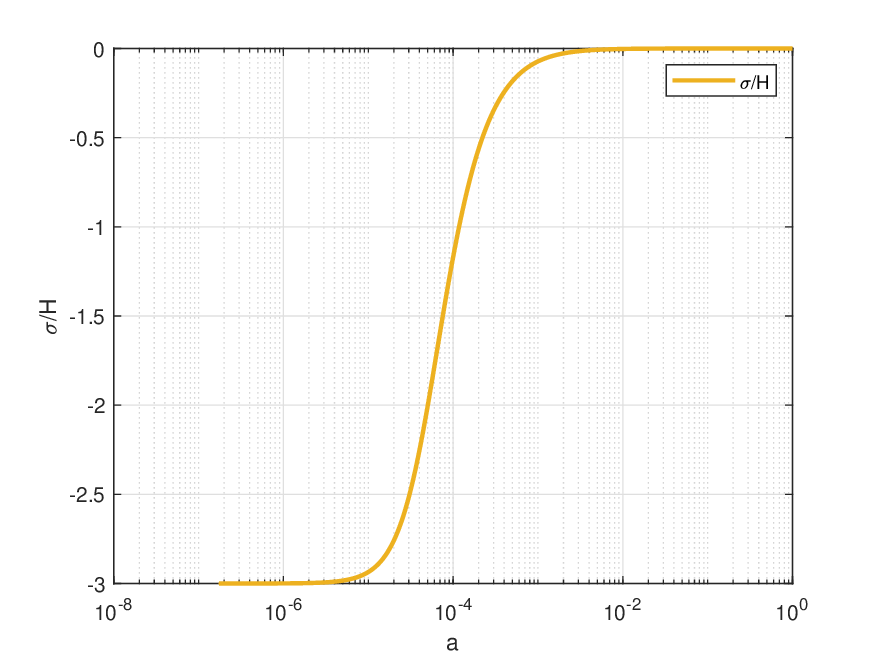}}\\
\subfloat[\label{fig:backaRhoSigLCDM}]{\includegraphics[width = 7.5 cm]{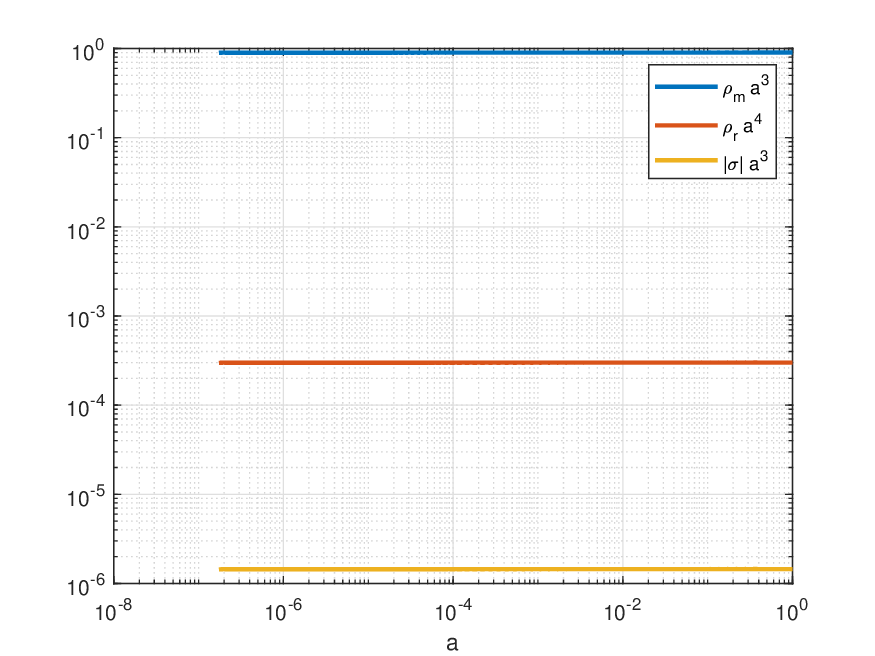}}\hfill
\subfloat[\label{fig:backOmegaLCDM}]{\includegraphics[width = 7.5 cm]{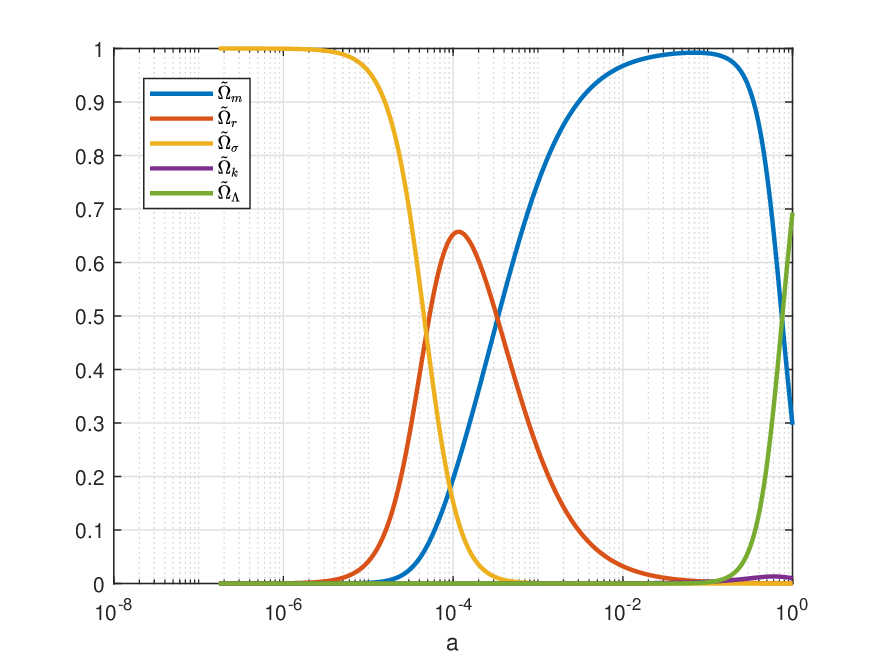}}
\caption{Evolution of dipole \lcdm\ with Planck \lcdm\ parameter values. {We evolve \eqref{dipole-LCDM-EoM} backward in time  with $a(0)=1, b(0)=0, \rho_r(0)=3\times 10^{-4}, \rho_m(0)=0.9, \Lambda=2.1, A_0=0.1$ and $\beta_m(0)=-10^{-7},\beta_r(0)=-5\times 10^{-4}$ at $t=0$. This data is consistent with Planck values $\Tilde{\Omega}_r\approx 10^{-4}, {\Omega}_\Lambda\approx 0.7, \Tilde{\Omega}_k\approx 0.01, \Tilde{\Omega}_m\approx 0.3$, together with $\Tilde{\Omega}_{\sigma}\sim 10^{-13}$.} Starting from $\beta_m,\beta_r\sim 0$ at the Big Bang, the relative tilt grows to $10^{-4}$, comparable to the CMB dipole. Evolution of relative densities are depicted in Fig.~\ref{fig:backOmegaLCDM}. For $a>10^{-3}$ this is essentially the same as usual Planck \lcdm\ except for the dipole, whereas at earlier times we have sizable shear contribution, yielding a later and less pronounced radiation dominance (compared to the usual Planck \lcdm\ case). }
\label{fig:back-Planck-LCDM}
\end{figure}

In much of the literature, see e.g. \cite{KMS, KMS-2, Coley-1, Coley-2, Coley-3}, the analysis has focused on late time behavior in tilted cosmology. However, in cosmology it is interesting to explore which initial conditions could have evolved to a given state now. In this section we study dynamics of dipole \lcdm\ backward in time to reach the initial singularity (if there is any). 
There are various choices of values of parameters today for the two fluid plus $\Lambda$ case, as well as the values of model parameter $A_0^2/\Lambda$. Here we discuss three representative classes of these initial conditions, depicted in Figs.~\ref{fig:back1}, \ref{fig:back2} and \ref{fig:back-Planck-LCDM}. 

\paragraph{Negative $\sigma$ case.} As we discuss below and depict in Fig.~\ref{fig:back1}, we explore trajectories with   $a(0)=1,b(0)=0$ at $t=0$ (present time) and $A_0=1,\Lambda=2.1 $. With these choices $\sigma$ is negative. These  values are just for illustrative purposes and not relevant to the values the Planck mission has reported for the \lcdm\ cosmology \cite{Planck-2018}; the latter will be discussed later in our third example. 
In this case if the sign of $\sigma$ does not change until reaching the $a=0$ singularity,  flows of both matter and radiation ($\beta_m,\beta_r$) go to zero as we approach the singularity. 

Fig.~\ref{fig:backa1} shows mild variations in $\rho_m a^3, \rho_r a^4, |\sigma| a^3$.  Fig.~\ref{fig:backsig1} shows that the shear $\sigma$ grows to to its minimum value $-3H$ and $\tilde\Omega_\sigma=1$ at early times -- the Big Bang is ``shear dominated'' as seen from Fig.~\ref{fig:backOmega1}.  Due to the presence of the tilts $\beta_m,\beta_r$ and recalling \eqref{sigma-rho-m-rho-r}, we see that the shear is sourced by the tilts. Therefore, the drop of $\sigma$ by the expansion is much slower than in the untilted ($\beta_m,\beta_r=0$) case. This is clearly seen in Figs.~\ref{fig:backsig1} and \ref{fig:backOmega1}, $\sigma/H$ essentially remains $-3$ for the first 10 e-folds ($e^{10}\sim 2\times 10^3$) and drops only by a factor of 3 in the next 2.3 e-folds ($\ln 10\sim 2.3$). This shows explicitly how Wald's cosmic no-hair theorem \cite{Wald-cosmic-no-hair} is affected by the tilt.

A large negative $\sigma$ can dominate the $-H$ term in the $\beta$-evolution equation  for $w=0$ case \eqref{p-m-dot-LCDM}, resulting in an initial phase of growing $\beta_m$. However, as  Fig.~\ref{fig:backbeta1} shows, when $\sigma=-3H/2$, $\beta_m$ reaches its extremum, in agreement with \eqref{p-m-dot-LCDM}, where the RHS of \eqref{p-m-dot-LCDM} vanishes and changes sign. For the radiation sector, we need to consider \eqref{beta-r-LCDM}. At early times the shear term dominates, acting as a positive source for $\beta_r$ and driving $\beta_r$ from zero to negative values. Then the last term in \eqref{beta-r-LCDM} is also added as a source for $\beta_r$ causing it to monotonically (but mildly) grow. It is notable that  both of the tilt parameters start from zero at the Big Bang, and the negative shear makes them grow. If one evolves $\beta_m, \beta_r$ further to the $a>1$ region (in future) as depicted in Fig.~\ref{fig:betaMMLCDML}, $\beta_m$ continues to go to zero very fast  while $\beta_r$ continues its mild growth.

As Fig.~\ref{fig:backa1} shows, $\rho_m a^3, \rho_r a^4, |\sigma| a^3$, start from constant asymptotic values at the Big Bang and stay there for several e-folds before dropping down at the present time. Of course, recalling the analysis and discussions in subsection \ref{sec:numeric}, one would expect these quantities to asymptote to constant values in the future too.

Finally, Fig.~\ref{fig:backOmega1} shows how Universe evolves from a shear-dominated Big Bang,  to  a brief  matter dominated epoch and finally a curvature dominated one. If we evolve further into future, we will find a $\Lambda$ dominated Universe as expected. 

\paragraph{Positive $\sigma$ case.} When $\beta_r,\beta_m$ are positive, then $\sigma$ is positive and they will not change sign. Fig.~\ref{fig:back2} shows evolution in such cases. Fig.~\ref{fig:backbeta2} shows that in accord with our analytic discussions,  $\beta_m, \beta_r >0$ at early times evolve to very small values at $a=1$ (today). $\sigma/H$ starts from sub-saturation values $\sigma/H\simeq 2.5$ and $\tilde\Omega_\sigma\sim 3/4$ at early times and monotonically decrease to zero, cf. Figs.~\ref{fig:backsig2} and \ref{fig:backOmega2}. This shows it is possible to have $|\sigma|/H\neq 3$ at the Big Bang.

Fig.~\ref{fig:backrho2} show that $\rho_r a^4, \rho_m a^3, \sigma a^3$ evolve to almost constant values today and Fig.~\ref{fig:backOmega2} show how the relative energy budget of the Universe among different components evolves: A shear-dominated Universe in few e-folds evolve to a radiation dominated Universe followed by a matter dominated epoch and then curvature-dominated era. Had we evolved further in future (in $a>1$ region), the curvature-dominated era would have been quickly replaced by a $\Lambda$ (dark energy) dominated era. A brief curvature dominated epoch is a result of our chosen values where $A_0\sim \sqrt{\Lambda}$.


\paragraph{Planck-\lcdm\ initial values.} We now set our input values (values of parameters today) in accord with Planck values \cite{Planck-2018} and explore the evolution of the Universe in the dipole \lcdm\ setting. To this end, we take $\Tilde{\Omega}_r\approx 10^{-4},{\Omega}_\Lambda\approx 0.7,\Tilde{\Omega}_k\approx 0.01,\Tilde{\Omega}_m\approx 0.3,\Tilde{\Omega}_{\sigma}\sim 10^{-13}$ at today $a=1$ and start with a smaller value for  $A_0$ and set $A_0=0.1$. Moreover, we assume $\beta_r=-5\times 10^{-4}$ at $a=1$ which is half of the rapidity of \rm{CMB} dipole and take $\beta_m=-10^{-7}$. We note that with our chosen units and values in Fig.~\ref{fig:back-Planck-LCDM} $H_0:=H(a=1)\simeq 1$.

We have chosen both $\beta_m, \beta_r$ to be negative and hence $\sigma<0$. In this respect evolution of tilts and the shear, shown in Fig.~\ref{fig:backbetaLCDM} and \ref{fig:backsigLCDM}, are qualitatively the same as those in Fig.~\ref{fig:back1}. In particular, note that very small negative values of $\beta_m,\beta_r$ evolves to a sizable \textit{relative} tilt between the matter and radiation sectors of the order CMB dipole value.

Fig.~\ref{fig:backaRhoSigLCDM} shows that $\rho_m a^3, \rho_r a^4, \sigma a^3$ are  almost constants. These have the same behaviour as the untilted (ie., the usual FLRW) Planck \lcdm\ case. This in particular means that contribution of the shear to the energy budget of the Universe goes as  $1/a^6$ which ultimately dominates in the very early Universe. In Fig.~\ref{fig:backOmegaLCDM} we have shown relative behaviour of  density parameters. A shear dominated Big Bang, after about 10 e-folds evolves to a radiation dominated Universe, followed by a matter dominated Universe in about 7 e-folds and a $\Lambda$ (dark energy) dominated era today. We note that in the radiation dominate epoch, radiation:shear:matter energy ratios are 65\%:17.5\%:17.5\%. The matter-DE equality happens in the last e-fold of the expansion of the Universe.  

An interesting feature of the dipole \lcdm\ evolution near the Big Bang (as we have seen in the examples discussed in this section) is that the Big Bang seems to shear dominated, a feature which seems fairly generic in the space of initial conditions. We note that this happens because shear is sourced by the tilts. Moreover, we note that we start with very small initial values of the tilt at the Big Bang. It is desirable to repeat this analysis with more generic initial values for the tilts and other parameters.

{The main message of this section is that by choosing parameters that are loosely consistent with those of Planck, we have been able to find evolutions in dipole cosmology that are loosely similar to those in the conventional flat \lcdm\ of FLRW. But despite this, we see that the relative tilt between matter and radiation can increase at late times suggesting that it is natural for at least some of the CMB dipole to be non-kinematical.}

\section{Discussion and Outlook}\label{sec:conclusion}

Our motivating goal in this paper was to illustrate that dipole cosmology is a viable model building paradigm, in exactly the same way that FLRW cosmology is. We can work with fluid mixtures and each component of the cosmic fluid can have its own flow, $\beta_i$. This observation is clearly of formal interest, but what makes it phenomenologically significant is that it is very natural that these Universes exhibit a {\it relative} flow between radiation and matter at late times. If the late time physics necessarily reduces to that of standard FLRW, dipole cosmology would be of limited interest as a paradigm that generalizes FLRW. Instead, what we have seen is that these cosmologies have an instability towards the generation of late time dipoles. Presence of this instability in FLRW cosmologies means that there are previously under-emphasized theoretical concerns in using FLRW as a setup for cosmological model building \cite{KMS-2}.  


For cosmic fluids with EoS $w>1/3$, this is a strong instability.\footnote{Since such stiff equations of state are not often used in cosmology, this should not be a big concern.} One of our key observations in this paper was that even for $w =1/3$, there exists an instability in the negative tilt ($A_0\beta<0$) cases. The growth of tilt for the radiation $w=1/3$ case is not sourced by the $(3 w-1) H$ term in \eqref{beta-growth-w-const}, because it identically vanishes. This yields milder growth 
of the tilt (compared to the $w>1/3$ cases) making it phenomenologically interesting for late time cosmology. This also dovetails with the fact that standard FLRW is a pretty good fit for the Universe as we know it. The tilt is sourced by a term that dies off exponentially at late times\footnote{Let us emphasize that the tilt itself grows (and saturates) -- this is a statement about the term sourcing the tilt.}, making it plausible that it may indeed be of physical significance in understanding late time dipoles. It is hence prudent to explore observational signatures of dipole \lcdm\ model and check if they could be associated with the dipole anomalies reported in the observational data, see \cite{Beyond-FLRW-review} for a detailed review. Note that many of the previous examples of tilt growth identified in single fluid models (eg. \cite{Coley-2,Coley-3, KMS-2}) are sourced by the $(3 w-1) H$ term, and apply only when $w>1/3$. 


Let us make a related observation. Even though we have suppressed it in this paper, we have in fact been able to find late time (negative) tilt growth sourced by these subleading terms {\it even in} cases where $w < 1/3$, when there is no cosmological constant. This has a simple analytical understanding: the $(3 w-1) H$ dies off at late times when there is no cosmological constant. This means that eventually the other terms on the RHS of \eqref{beta-growth-w-const} will dominate, leading to tilt growth. If there is a cosmological constant, the $(3 w-1) H$ term asymptotes to a negative constant and will eventually come to dominate the dynamics and will kill the flow. This observation seems to have been missed in previous work. We have been able to explicitly demonstrate this expectation numerically up to $w$ as low as $ \sim 1/5$, but we expect that with suitable initial conditions the qualitative results should remain identical all the way to $w > 0$. 
We have not discussed these results in detail in the present paper, because presumably $w=1/3$ and $w=0$ are more interesting for phenomenology. Nonetheless, this observation does show that homogeneous tilt instability is even more generic than what we argued for in \cite{KMS-2}.

In previous papers \cite{KMS, KMS-2}, we have noted that dipole cosmology has an instability towards late time tilt growth in specific cases, including a large subclass of cases where we had a total equation of state $w(t) \rightarrow -1$ at late times. This made us suggest that the cosmological principle has an instability towards tilt growth at late times -- if not universally, at least quite generically in physically interesting settings. The results of this paper show that this is in fact true even for very conventional models involving only radiation and matter. These models have an instability towards the growth of a  homogeneous flow, a dipole, at late times. We explicitly demonstrated this in this paper, in a natural generalization of the \lcdm\ model to the dipole setting, a model that we called dipole \lcdm\ model. The observation is, however, far more general. All one has to do to see these instabilities is that the radiation starts off with a flow that is negative (in our conventions, where $A_0=+1$).

In this paper, we have assumed that there is a unique flow velocity in the pressureless matter sector of dipole \lcdm\ model. Even though we have not discussed it in detail here, it is straightforward (and  natural) to allow the possibility that the fluid flows for dark matter and baryonic matter (even though both have $w=0$) are distinct. This is equally straightforward to explore at the background level, but since in order to make a fully realistic comparison with data we need to also consider perturbations, we have suppressed it in this paper. To illustrate the key observation that radiation and matter can have a growing relative dipole at late times, these details were not necessary. But to construct a potentially fully realistic model, we will need to incorporate such nuances. A clear open problem for the future is to consider perturbation theory in the dipole cosmology setting, if we are hopeful of connecting things to data.

Finally, let us conclude with some comments about the significance of these observations for late time tensions in cosmology. Our result shows that there is a late time instability in the standard FLRW model, which can be of phenomenological relevance for cosmic dipoles. Whether these observations can fully resolve the cosmic tensions remains to be seen. It has been noted in \cite{H0-HSA, H0-HSA-2} that dipole anisotropies in $H_0$ can at most explain a $\Delta H_0 \sim 4-5 $ km/s/MpC in the Hubble tension. So at least naively, these anisotropies are insufficient to explain away the tensions between CMB values and late time values of $H_0$ \cite{Crisis-1, Crisis-2}. However, to systematically make an argument of this type, one has to redo the phenomenology and data analysis in the dipole cosmology setting from scratch. Only after that, can we systematically evaluate the status of late time tensions in dipole cosmology. In any event, it is clearly of interest to further study the late time instability of the FLRW paradigm that we have uncovered in this paper, given the fact that late time Universe is ripe with puzzles.  

\section*{Acknowledgments}

We thank Alireza Allahyari, Justin David and Prasad Hegde for discussions. The work of EE and MMShJ is supported in part by SarAmadan grant No ISEF/M/401332. 

\appendix

\section{Multiple Flows}\label{Appendix:consistency}

In this appendix we discuss further the fact that  dipole cosmology systems allow for distinct fluid components, each with their own pressure, density and tilt (as long as the flows are all along the same direction, which we call $z$ in our notation). The argument works the same way for any number of fluids. Our discussion is in the metric language, but it can also be phrased in terms of vierbeins and spin connections which is the setting of \cite{King}. Multiple fluids in the vierbein approach will be discussed in \cite{KM-upcoming} in a setting that is slightly more general than the $U(1)$ isotropic case that is the purview of our discussion in the present paper. 

Our starting point is the Einstein equations \eqref{cosmoC}, where the total energy-momentum tensor in the RHS of the equation is given in \eqref{T-sum}. For an $n$ component fluid, 
\begin{equation}
T^\mu{}_\nu=\sum_i (T^\mu{}_\nu)_i,
\end{equation}
where
\begin{equation}
\small{
    (T_{\mu\nu})_i = \left(
\begin{array}{cccc}
 {\rho_{i} +(\rho_{i}+p_{i}) \sinh ^2\beta_{i}} & -\frac12(\rho_{i} +p_{i}) X \sinh 2\beta_{i}  & 0 & 0 \\
 -\frac12(\rho_{i} +p_{i}) X \sinh 2\beta_{i}  & \left(p_i+(\rho_{i} +p_{i}) \sinh ^2\beta_{i} \right) X^2 & 0 & 0 \\
 0 & 0 & p_{i}\ e^{-2 A_{0} {z}}\ Y^2 & 0 \\
 0 & 0 & 0 & p_{i}\ e^{-2  A_{0} {z}}\ Y^2 \\
\end{array}
\right)}
\end{equation}
and
\begin{equation}
X(t)= a e^{2b}, \qquad Y(t)=a e^{-b}.
\end{equation}
Consistency of Einstein equations as usual implies $\nabla_\mu T^\mu{}_\nu=0$, which takes the explicit form
\begin{eqnarray}
\sum_{i }\cosh^{2}\beta_{i}\Big(\dot{\rho}_{i}+(\rho_{i}+p_{i})(\frac{\dot{X}}{X}+\frac{2\dot{Y}}{Y}+\tanh{\beta_{i}}\dot{\beta}_{i}-\frac{2A_0\tanh{\beta_{i}}}{X})\Big) + \hspace{1in} \nonumber \\
 + \sinh^{2}\beta_{i}\Big(\dot{p}_{i}+(\rho_{i}+p_{i})(\frac{\dot{X}}{X}+\coth{\beta_{i}}\dot{\beta}_{i})\Big) = 0 \hspace{0.2in} \label{1stT}\\
\sum_{i } \sinh{2\beta_{i}}\Big(\dot{\rho}_{i}+\dot{p}_{i} + (\rho_{i}+p_{i})(\frac{2\dot{X}}{X} + \frac{2 \dot{Y}}{Y}) + \tanh{\beta_{i}}\dot{\beta}_{i} + \coth{\beta_{i}}\dot{\beta}_{i}-\frac{2A_0\tanh{\beta_{i}}}{X}\Big) = 0  \label{2ndT}
\end{eqnarray}

Covariant-constancy of energy-momentum tensor for {\em each} component of the cosmic fluid $(T^\mu{}_\nu)_i$ is needed to have an autonomous system of equations. This follows from the assumption that fluid components do not interact with each other, while they all covariantly couple to the background metric. Upon this assumption, we then have 
\begin{equation}\label{Ti-constancy}
   \nabla_\mu (T^\mu{}_\nu)_i=0, \qquad \forall \ i. 
\end{equation}
Recalling \eqref{T-tilted}, and allowing for each component to have its own tilt parameter $\beta_i$, the above yields,
\begin{subequations}\label{rho-p-multifluid}
\begin{align}
\dot{\rho}_i+3H(\rho_i+p_i)&=-(\rho_i+p_i)\tanh\beta_i(\dot{\beta}_i-\frac{2A_0}{a} e^{-2b}) \label{Con-rho-i} \\
\dot{p}_i+H(\rho_i+p_i)&= -(\rho_i+p_i)\left( \frac23\sigma+\dot{\beta}_i\coth{\beta}_i\right). \label{Con-p-i}
\end{align}
\end{subequations}
for all $i$. It can be checked that these equations for the distinct components can be algebraically combined to give precisely \eqref{1stT} and \eqref{2ndT}. 

Einstein equations yield two equations which govern evolution of $a,b$ (or $H,\sigma$). To be more precise, there are two equations which involve $H, \sigma$ and not their time derivatives, and two equations that involve $\dot{H}, \dot{\sigma}$ (which can be traded for the total covariant conservation laws). The two first derivative equations are,
\begin{subequations}\label{EoM-H-sigma-multifluid}
\begin{align}
H^2-\frac19\sigma^2-\frac{A_0^2}{a^2} e^{-4b}&=\sum_i\frac{\rho_i}{3}+\frac13 (\rho_i+p_i)\sinh^2\beta_i
\label{EoM-H-sigma-sum}\\
 \sigma &=\frac{1}{4A_0} a e^{2b}\ \sum_i(\rho_i+p_i)\sinh2\beta_i.
\label{EoM-sigma-sum}
\end{align}
\end{subequations}
As in the  usual FLRW case, one may define modified density parameters
\begin{equation}\label{Omega's}
   3H^2 \tilde{\Omega}_{i}:=\rho_i+(\rho_i+p_i)\sinh^2\beta_i, \qquad H^2\tilde{\Omega}_k:=\frac{A_0^2}{a^2e^{4b}},\qquad \tilde{\Omega}_{\sigma}:=\frac{\sigma^2}{9H^2}
\end{equation}
where $\tilde\Omega$'s are positive variables. In terms of the above \eqref{EoM-H-sigma-sum} takes the simple ``energy budget'' form
\begin{equation}\label{sum-Omega=1}
    \sum_i \tilde\Omega_i +\tilde\Omega_k+\tilde\Omega_\sigma=1. 
\end{equation}
We note that if our fluids satisfy weak energy condition $(\rho_i \geq 0, \rho_i+p_i \geq 0)$, $\tilde\Omega$'s are all in $[0,1]$ range. 
Note also that cosmological constant $\rho=\Lambda, p=-\Lambda$, is a particular case which allows for constant $\rho,p$ and arbitrary $\beta$, as \eqref{rho-p-multifluid} shows. For this case also, one can define 
\begin{equation}
 \Omega_\Lambda:=\frac{\Lambda}{3H^2}   
\end{equation}
which can be viewed as one of $\tilde\Omega_i$'s.  It may happen that in the course of evolution of the Universe in specific epochs one of the $\tilde\Omega$'s dominate the sum, i.e. for one of the $\tilde\Omega_i$'s, $\tilde\Omega_i\simeq 1$ while the others are negligible. 

For a generic $n$ component cosmic fluid plus a cosmological constant $\Lambda$, \eqref{rho-p-multifluid} and \eqref{EoM-H-sigma-multifluid} give $2n+2$ equations for  $3n+2$ variables, $\rho_i, p_i,\beta_i; a, b$. As usual, these equations becomes autonomous once we supplement them with EoS, $p_i=w_i \rho_i$ for a given $w_i$. For constant $w_i$ cases we get
\begin{subequations}\label{const-w-dipole-multifluid}
\begin{align}
&\rho_i^{\frac{w_i}{1+w_i}} a e^{2b} \sinh\beta_i =  C_i, \qquad \forall i, 
\label{const-w-dipole-rho-X-beta-multifluid}\\
&\dot{\beta}_i\big(\coth\beta_i-w_i \tanh \beta_i\big)=(3w_i-1)H-\frac23\sigma-\frac{2w_i A_0}{a(t)} e^{-2b}\tanh\beta_i \quad \forall i \label{beta-growth-w-const-multifluid}
\\
&H^2=\frac13\Lambda+\frac{A_0^2}{a^2} e^{-4b}+  \sum_i \frac{\rho_i}{3}[1+(1+w_i)\sinh^2\beta_i]+\frac19\sigma^2 
\label{EoM-H-multifluid}\\
&\sigma =\frac{1}{4A_0} a e^{2b}\ \sum_i \rho_i(1+w_i)\sinh2\beta_i=\frac{1}{2A_0} \sum_i C_i \rho_i^{\frac{1}{1+w_i}} \cosh\beta_i,
\label{sigma-w-dipole-multifluid}
\end{align}
\end{subequations}
where $C_i$ are integration constants. In the above we explicitly considered a cosmological constant and assumed $w_i\neq -1$ for the rest of cosmic fluid components. One may describe the system through density parameter variables \eqref{Omega's}. These variables explicitly solve \eqref{EoM-H-multifluid} and we have $2n$ variables $\tilde\Omega_i, \beta_i$ describing the matter sector and $\tilde\Omega_\sigma, \tilde\Omega_k$ the metric sector. Eqs.~\eqref{beta-growth-w-const-multifluid}, \eqref{EoM-H-sigma-multifluid} and  \eqref{sigma-w-dipole-multifluid} may be written as
\begin{subequations}
\begin{align}
\sum_i \tilde\Omega_i &+\Omega_\Lambda+\tilde\Omega_\sigma+ \tilde\Omega_k=1, \\
\frac{\dot{\beta}_i}{H}\big(\coth\beta_i-w_i \tanh \beta_i\big)&=(3w_i-1)-2\ \text{sgn}(\sigma)\sqrt{\tilde\Omega_\sigma}-2w_i\sqrt{\tilde\Omega_k}\tanh\beta_i \qquad \forall i \label{beta/H-growth-w-const-multifluid}
\\
4\ \text{sgn}(\sigma)\sqrt{\tilde\Omega_\sigma\, \tilde\Omega_k} &=\sum_i \ \tilde\Omega_i\ \frac{(1+w_i)\sinh2\beta_i}{1+(1+w_i)\sinh^2\beta_i}.     \label{Omega-i-Omega-k}
\end{align}
\end{subequations}
where $\text{sgn}(\sigma)$ is the sign of $\sigma$. Note that $\sigma$ can be positive and negative while  $A_0$ is  positive by assumption. We note also that each term in the sum in the RHS of \eqref{Omega-i-Omega-k} can be positive or negative depending on the sign of $\beta_i$.

\end{document}